\documentclass[12pt]{article}
\usepackage{paper2e}
\usepackage{mydefs2e}
\usepackage{epsf}

\newcommand{\ms}{M_{\rm SUSY}}

\def\caption#1{{\centerline{\vbox{\baselineskip=14pt
        \vskip.15in\hsize=5.5in\noindent{#1}\vskip.1in }}}}

\begin{document}
\begin{titlepage}

\preprint{UMD-PP-99-102}

\begin{center}
\Large\bf
Realistic Anomaly-mediated\\\medskip
Supersymmetry Breaking
\end{center}

\author{Z. Chacko,%
\footnote{E-mail: {\tt zchacko@bouchet.physics.umd.edu}}
\ 
Markus A. Luty,%
\footnote{Sloan Fellow.
E-mail: {\tt mluty@physics.umd.edu}}
\ 
Ivan Maksymyk%
\footnote{E-mail: {\tt maksymyk@physics.umd.edu}}
\ 
Eduardo Pont\'on%
\footnote{E-mail: {\tt eponton@wam.umd.edu}}}

\smallskip
\address{Department of Physics\\
University of Maryland\\
College Park, Maryland 20742, USA}

\begin{abstract}
We consider supersymmetry breaking communicated entirely
by the superconformal anomaly in supergravity.
This scenario is naturally realized if supersymmetry is broken
in a hidden sector whose couplings to the observable sector
are suppressed by \emph{more} than powers of the Planck scale,
as occurs if supersymmetry is broken in a
parallel universe living in extra dimensions.
This scenario is extremely predictive:
soft supersymmetry breaking couplings
are completely determined by anomalous dimensions
in the effective theory at the weak scale.
Gaugino and scalar masses are naturally of the same order,
and flavor-changing neutral currents are automatically suppressed.
The most glaring problem with this scenario is that
slepton masses are negative in the minimal supersymmetric
standard model.
We point out that this problem can be simply solved by
coupling extra Higgs doublets to the leptons.
% Unification can be preserved by adding triplets, and 
Lepton
flavor-changing neutral currents can be naturally avoided by
approximate symmetries.  % or selection rules.
We also describe more speculative solutions involving
compositeness near the weak scale.
% Realistic electroweak symmetry breaking cannot be obtained by
% adding an explicit $\mu$ term, so
We then turn to electroweak symmetry breaking.
Adding an explicit $\mu$ term gives a value for $B\mu$ that is
too large by a factor of $\sim 100$.
We construct a realistic model in which the $\mu$ term arises from the
vacuum expectation value of a singlet field, so \emph{all}
weak-scale masses are directly related to $m_{3/2}$.
We show that fully realistic electroweak symmetry breaking can
occur in this model with moderate fine-tuning.
\end{abstract}

%\date{April, 1999}

\end{titlepage}

% ----------------------------------------------------------------------------
\section{Introduction}
% ----------------------------------------------------------------------------
% Supersymmetry (SUSY) is arguably the most attractive framework for resolving
% the hierarchy problem, and supergravity (SUGRA) is an extremely natural
% messenger for SUSY breaking.
The hidden-sector scenario for supersymmetry (SUSY) breaking is arguably
the simplest and most natural mechanism for realizing SUSY in nature
\cite{Hidden}.
In this scenario, one assumes that SUSY is broken in a hidden sector that
couples only gravitationally to observable fields.
Because supergravity (SUGRA) couples universally to all fields, it necessarily
connects the observable and hidden sectors, and therefore communicates
SUSY breaking to the observable sector.
% If there are no stronger interactions that connect the hidden and
% observable sectors, SUGRA mediation will be the leading effect.
This scenario is very attractive from a theoretical point of view because
all of the ingredients are either there of necessity (\eg SUGRA) or
arise naturally (\eg hidden sectors are a generic consequence of string
theory).

SUGRA interactions are flavor-blind, so one might hope that this scenario
will not give rise to off-diagonal terms in the squark masses that can
lead to flavor-changing neutral currents.
Unfortunately, it is very difficult to suppress higher-dimension
operators of the form
\beq\eql{badop}
\scr{L}_{\rm eff} \sim \myint d^4\th\,
\frac{1}{m_*^2} X^\dagger X Q^\dagger Q,
\eeq
where $X$ is a field in the hidden sector, $Q$ is a field in the
observable sector, and $m_* \sim 2 \times 10^{18}\GeV$
is the reduced Planck mass.
If SUSY is broken by $\avg{F_X} \ne 0$, this will give rise to soft
masses $m_{\tilde{Q}} \sim \avg{F_X} / m_* \sim m_{3/2}$.
The difficulty is that there is {\it a priori} no reason that the terms
\Eq{badop} (and hence the scalar masses) should be flavor-diagonal,
and this will lead to unacceptable flavor-changing neutral currents
(FCNC's) unless \cite{SUSYFCNC}
\beq
\frac{m^2_{\tilde{d}\tilde{s}}}{m^2_{\tilde{s}}}
\lsim (5 \times 10^{-3}) \left( \frac{m_{\tilde{s}}}{1\TeV} \right),
\qquad
\Im \left(\frac{m^2_{\tilde{d}\tilde{s}}}{m^2_{\tilde{s}}} \right)
\lsim (5 \times 10^{-4}) \left( \frac{m_{\tilde{s}}}{1\TeV} \right).
\eeq
We know that SUGRA is an effective theory with a cutoff of order $m_*$,
and we expect that the fundamental theory above this cutoff does not
conserve flavor.%
\footnote{One can consider models
in which the flavor symmetry is a gauged symmetry at the Planck
scale.
However, flavor symmetry must be broken to obtain the observed quark
and lepton masses, and it is nontrivial to do this while maintaining
sufficient degeneracy among squarks to avoid FCNC's \cite{SUSYflavor}.}
The operator \Eq{badop} is therefore allowed by all symmetries, and
there seems to be no natural way to suppress it.
This is the famous SUSY flavor problem.

An elegant solution to this problem was proposed by Randall and
Sundrum in \Ref{RS}.
They considered a scenario in which there are one or more compact
extra dimensions with a size $R \gg 1/M_*$, where $M_*$ is the fundamental
(higher-dimensional) Planck scale.
They further assumed that the observable
sector and SUSY breaking sector are localized on separate
$(3 + 1)$-dimensional subspaces in the extra dimensions.
These subspaces may arise as topological defects in field theory, or
D-branes or orbifold fixed points in string theory;
their precise nature is not essential for the present purpose, and
we will refer to them generically as `3-branes'.
For simplicity, we assume that the separation between the branes
in the extra dimensions is also of order $R$.
We also assume that the only fields with mass below $M_*$ that
propagate in the higher dimensions between the branes are the
SUGRA fields
(which are necessarily present, since gravity is the dynamics of
spacetime itself).

Scenarios with SUSY breaking localized on orbifold fixed points
have been considered
previously by Ho\v rava and Witten in the context of M theory \cite{HW}.
There the existence of a large compactification radius reduces the
fundamental string scale and allows gauge-gravity unification.
String vacua with gauge and/or matter fields localized on D-branes
have been considered by several authors \cite{Dvac}.
The observation of \Ref{RS} that the SUSY flavor problem can be
solved in an elegant way in models of this type gives a strong
additional motivation for these scenarios.

The key observation of \Ref{RS} is that in this scenario operators
such as \Eq{badop} are not present in the effective theory below
$M_*$ for the excellent reason that they are not local operators!
Even if we assume that the theory above the scale $M_*$ violates
flavor maximally, the only flavor-violating couplings between the
observable and hidden sectors comes from the exchange of quanta
with Planck-scale masses between the branes.
But these effects are exponentially suppressed by $e^{-M_* R}$
due to the spatial fall-off of massive propagators.
The leading coupling between the hidden and observable sectors
comes from SUGRA fields, whose couplings are flavor-blind.
% In this sense, this scenario realizes the original hope of
% hidden-sector SUSY breaking.
Note also that due to the exponential suppression of flavor-changing
effects, this scenario requires only a very modest hierarchy between
$R$ and $M_*$.

Below energy scales of order $1/R$, the effective theory becomes
$(3 + 1)$-dimensional, with the hidden and observable sectors
inhabiting the same space.
This effective lagrangian contains higher-dimension operators
connecting the hidden and observable sectors from integrating
out Kaluza-Klein modes, but these effects
conserve flavor and therefore cannot contribute to FCNC's.
Furthermore, these effects do not
compete with anomaly mediation unless $R \sim 1/M_*$,
as discussed below.

% and exchange of these quanta
% can give rise to higher-dimension operators connecting the hidden
% and observable sectors.
% However, these contact terms are suppressed both by powers of
% $M_{\rm KK}$ (from the heavy masses) \emph{and} $M_*$ (from the
% gravitational strength coupling.

If we assume that the observable sector contains only renormalizable
couplings, tree-level SUGRA effects do not give rise to soft SUSY
breaking in the observable sector.
At loop level, soft SUSY breaking is generated, in a way that is
connected in a precise way to the conformal anomaly \cite{RS,GLMR}.
% (This will be reviewed below.)
This leads to an extremely predictive scenario:
\emph{all} of the soft SUSY breaking parameters are
determined by $m_{3/2}$ and anomalous dimensions at the electroweak
scale.
Specifically,
\beq\eql{exactsoft}
m_\la = \frac{\be(g^2)}{2 g^2} m_{3/2},
\qquad
m^2_{\tilde{Q}} = -\frac{1}{4}
\frac{d \ga_Q}{d \ln\mu} m_{3/2}^2,
\eeq
where
\beq
\be(g^2) \equiv \frac{d g^2}{d\ln\mu},
\qquad
\ga_Q \equiv \frac{d \ln Z_Q}{d\ln\mu}
\eeq
are the gauge beta function and matter field
anomalous dimensions.
\Eqs{exactsoft} are \emph{exact} formulas for the SUSY breaking
parameters in a superfield coupling
scheme \cite{ALR,GLMR} if we ignore quantum gravity corrections.
An important feature of these formulas is that
scalar and gaugino masses are of the same order
\beq\eql{msoft}
m_{\la} \sim m_{\tilde{Q}} \sim \ms = \frac{m_{3/2}}{16\pi^2}.
\eeq
These results hold in any scenario with additional suppressions between
the hidden and observable sectors.
We therefore refer to this scenario as `anomaly-mediated SUSY breaking.'

The quark anomalous dimensions (and hence the squark masses)
are dominated by the contribution of
$SU(3)_C$, which is flavor-independent.
The only flavor-dependent contributions come from the quark Yukawa
couplings, which are small for the first two generations, so FCNC's
are suppressed.%
\footnote{
A detailed study of FCNC's for the third generation in this scenario
would be worthwhile.}
% Also, FCNC's are naturally suppressed because the squark masses are
% (to a very good approximation) diagonal in the mass eigenbasis.

Anomaly mediation is clearly an attractive scenario, but
in its simplest form, it is not realistic.
The most glaring problem is that if the observable
sector consists of the minimal supersymmetric standard model
(MSSM), all slepton mass-squared terms are negative, leading to a
spontaneous breaking of electromagnetism.
\Ref{RS} considered the possibility that this is cured by having
additional interactions in the bulk coupling the leptons to the
hidden sector.
However, these new contributions depend on additional parameters
that must be adjusted to special values in order to obtain soft
masses of order \Eq{msoft}.
Therefore, in such scenarios the fundamental requirement that all
soft masses are of the same order does not arise naturally.
% Also, putting additional fields in the bulk raises anew the specter
% of FCNC's, and presumably complicates the task of engineering a
% realistic model.
In this paper, we explore the alternative that all SUSY breaking
arises from anomaly mediation, so that \Eq{msoft} is automatic.
This requires an extension of the MSSM.
In order to solve the problem of slepton masses, at least some of the
fields beyond the MSSM must have masses near the weak scale.
This is because the slepton masses are
determined by the anomalous dimensions at the weak scale.
As a result, this kind of model is directly testable in
accelerator experiments, giving an additional motivation to study such
models.

We consider several possibilities for obtaining positive slepton
mass-squared terms.
One simple possibility is to add an
extra pair of Higgs doublets with large (order 1) Yukawa couplings
to leptons.
Lepton FCNC's can be naturally suppressed by approximate flavor
symmetries if we introduce 3 extra pairs of Higgs doublets.
In order to preserve
one-step gauge unification, one can contemplate adding 3
color triplets, so that the extra fields form complete
$({\bf 5} \oplus \bar{\bf 5})$'s of $SU(5)$.
However, this makes the $SU(3)_C$ beta function vanish at one loop,
leading to squark masses that are too small.
We therefore give up one-step gauge coupling in this approach,
although of course models of this type may be embedded in a non-minimal
grand unified theory, \eg with intermediate scales.

We also briefly discuss
a more speculative mechanism for positive slepton masses
involving compositeness at the
weak scale.

We then turn to electroweak symmetry breaking.
Introducing a tree-level $\mu$ term spoils the relation
\Eq{msoft}, which is required for a realistic model.
This is a direct reult of introducing a
dimensionful parameter into the theory.
(Specifically, $B\mu = m_{3/2} \mu \sim 16\pi^2 \mu\ms$.)
The Giudice-Masiero mechanism \cite{GM} for generating a $\mu$ term
is not available if we do not want to introduce additional couplings
between the hidden and observable sectors.%
\footnote{Even if we do introduce such interactions, special
parameter choices are required
to obtain $B\mu \sim \mu\ms$ \cite{RS}.}
The remaining possibility is that an effective $\mu$ term is
generated by the vacuum expectation value (VEV) of a singlet
at the weak scale.
Motivated by the solution of the slepton mass problem, we consider
a model with 3 extra Higgs doublets, 1 vector-like pair of color triplets,
and 4 singlets.
The color triplets are needed in order to obtain a negative mass-squared
for the singlet whose VEV gives the $\mu$ term.
The other
singlets give important contributions to the soft terms for the
ordinary Higgs fields required for electroweak symmetry breaking.
The model contains no dimensionful parameters, so \emph{all} mass scales
are set by anomaly mediation:
in this sense, this model takes the idea of radiative symmetry
breaking to its logical extreme.
It is remarkable that the masses of all superpartners can be given
phenomenologically acceptable values through this mechanism,
with only moderate fine-tuning ($\sim 1/20$).

This paper is organized as follows.
In Section 2, we review SUSY breaking on a parallel
universe and anomaly mediation.
In Section 3, we present our solutions to the slepton mass problem.
In Section 4, we show how electroweak symmetry breaking (including the
$\mu$ term) can arise entirely from anomaly mediation.
Section 5 contains our conclusions.

% ----------------------------------------------------------------------------
\section{Review of Anomaly-Mediated Supersymmetry Breaking}
% ----------------------------------------------------------------------------
In this Section, we give a brief review of anomaly-mediated SUSY
breaking.
No originality is claimed here, but we hope that an overview
will be useful to the reader.
We also wish to emphasize the simple `1PI' understanding of anomaly
mediation described in \Ref{GLMR}.

% ----------------------------------------------------------------------------
\subsection{Parallel Universes}
Anomaly-mediated SUSY breaking is the leading effect in models where
% the hidden sector is `sequestered' in the sense that
higher-dimension operators connecting the hidden and observable sectors
have coefficients that are small in units of the Planck mass.
\Ref{RS} pointed out that this scenario
is naturally obtained if the hidden sector is localized on a parallel
universe in extra dimensions.
% (They refer to the hidden sector as `sequestered' is this scenario.)
Although anomaly-mediation is in principle more general than this
scenario, we briefly review some of the most important features of
SUSY breaking communicated from a spatially separated `3-brane'
to see how some of the general features discussed below can arise
in a specific context.

Our starting point is the assumption that there are $n \ge 1$
extra dimensions compactified with a radius 
$R \gg 1/M_*$, where $M_*$ is the
$(4 + n)$-dimensional Planck scale.
% Most of our estimate are independent of the details of the
% geometry of the compactified space as long as all dimensions are of order
% $R$.
When we perform numerical estimates we will want to include \eg factors
of $\pi$, and for simplicity we will take the compactified space to be a
symmetric torus with radius $R$.
% (Later we will see that $M R \sim 10$ is sufficient.)
Furthermore, we assume that the standard-model fields are localized
on a $(3 + 1)$-dimensional subspace (`3-brane'), and the hidden sector
that breaks SUSY is localized on a 3-brane spatially separated from
the observable 3-brane by a distance $\sim \pi R$.
(This is the maximum distance between two points on a circle of radius
$R$, and the most natural choice for the separation of the 3-branes.)
We also assume that the only light (below $M_*$) fields propagating
between the branes are SUGRA fields.
This is a very strong assumption, and we will briefly
consider the extent to which it can be relaxed below.
We will not address the question of how such a scenario can arise from
Planck-scale physics, but we note that extra dimensions and branes
(in our generalized sense) with localized degrees of freedom are generic
features of string vacua.
We will simply assume that such a configuration exists (and is stable)
and work out the consequences.

We do this by writing an effective theory below the scale $M_*$ that
includes the branes, the fields localized on them, and SUGRA.
The effective action for such a theory is
\beq
S^{(4 + n)}_{\rm eff} = \myint d^4 x\, d^n y\, \Bigl\{
\de^n(y - y_{\rm obs}) \scr{L}^{(4)}_{\rm obs}
+ \de^n(y - y_{\rm hid}) \scr{L}^{(4)}_{\rm hid}
+ \scr{L}^{(4 + n)}_{\rm bulk} \Bigr\},
\eeq
where $y$ are the coordinates corresponding to the extra dimensions.
(We are
considering the simple case where the hidden and observable sectors
are localized at fixed values of $y$.)
We do not make any assumptions about the symmetry structure of 
the theory
above the scale $M_*$, and so we include all higher-dimension operators
consistent with gauge symmetries.
However, there are no higher-dimension operators
connecting the fields in the hidden and observable sectors because
such operators are not local.
In fact, if we integrate out heavy modes with masses of order $M_*$
that propagate between the hidden and observable sectors, these will
give effects suppressed by $e^{-\pi R M_*}$ because of the exponential
decay of a massive propagator in position space (in any dimension).%
\footnote{As discussed above, these exponentially suppressed effects cannot be
parameterized by local terms in the effective field theory below $M_*$.
This would seem to imply that there is no way to consistently
include these effects in an effective field theory description.
We believe that the resolution to this apparent paradox is that
effective field theory is capable of reproducing the results of the
full theory \emph{in an expansion in $1/M_*$}.
The exponential effects are smaller than any power of $1/M_*$, and hence
fall outside the domain of effective field theory.}
See Fig.~1a.
These effects are exponentially suppressed if $R \gg 1/M_*$;
in fact, we need only a modest hierarchy {\eg\ $M_* R \sim 3$) in
order to be completely safe from FCNC's induced by higher-dimension
operators.%
\footnote{This does not forbid operators that violate \eg\ baryon
number within the visible sector.
However, such operators can be suppressed in other ways, such as
imposing extra gauge symmetries.}
The fact that very modest hierarchies of scales are required is
one of the most appealing features of this scenario.

\begin{figure}[t]
\centerline{\epsfxsize=5.0in\epsfbox{
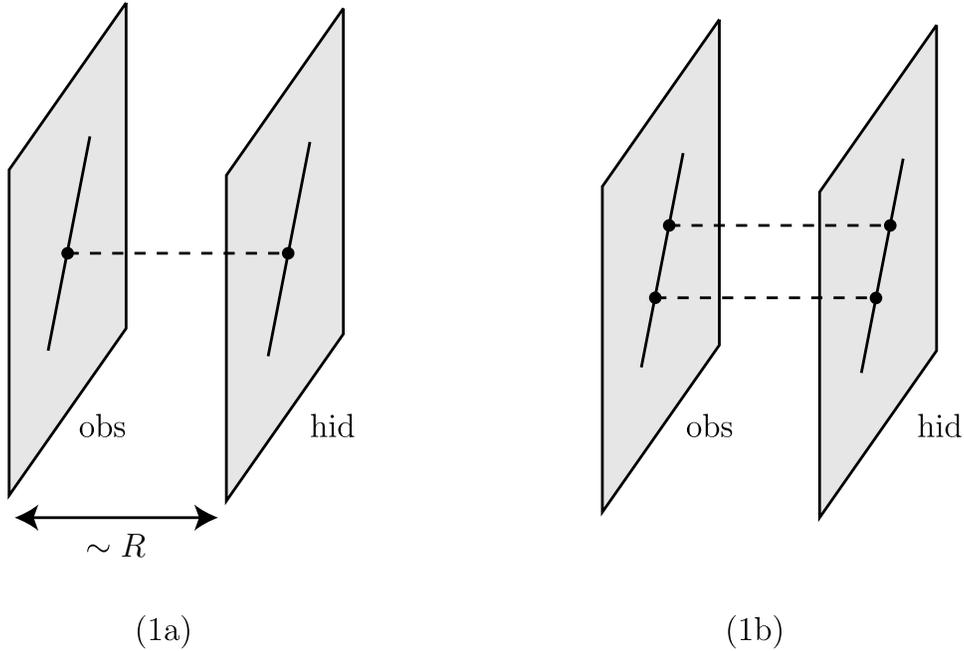}}
\smallskip
\caption{Fig.~1.\ \ Diagrams that contribute to quartic terms in
the \Kahler potential connecting the hidden and observable sectors.
The solid lines correspond to fields localized on the 3-branes,
while the dashed lines correspond to SUGRA fields propagating in
the bulk.}
\end{figure}

We can now construct the effective theory below the scale $R^{-1}$.
This can be viewed as integrating out all the Kaluza-Klein (KK) excitations
of the bulk SUGRA modes with mass $M_{\rm KK} \sim R^{-1}$.
The resulting theory is a 4-dimensional effective SUGRA theory with a
a cutoff $\La \sim M_{\rm KK}$ and a 4-dimensional Planck scale given by
\beq\eql{mPlanck}
m_*^2 = V_n M_*^{2 + n},
\eeq
where $V_n$ is the volume of the extra dimensions.
($V_n = (2\pi R)^n$ for a symmetric torus.)
An important point is that this matching involves integrating out only
SUGRA modes, and therefore does not induce flavor-dependent operators.
However, we must estimate these matching effects to see whether they
give significant contributions to flavor-diagonal soft terms.
Exchange of a single SUGRA KK mode (as in Fig.~1a) is suppressed by
$1/(m_*^2 M_{\rm KK}^2)$, with the 4-dimensional Planck suppression
arising from the gravitation coupling and $M_{\rm KK}$ from the massive
propagator.
These diagrams therefore give higher-derivative operators such as
\beq
\De\scr{L}^{(4)} \sim \frac{1}{m_*^2 M_{\rm KK}^2}
\myint d^4\th\, X^\dagger X
Q^\dagger \Yfund Q 
\eeq
that do not contribute to soft masses.
There are also loop effects connecting the
hidden and observable sectors, as in Fig.~1b.
% If the quanta being exchanged are KK modes, this diagram
% is suppressed by $1/(M_{\rm KK}^4 m_*^4)$, which can be neglected.
If the quanta being exchanged are massless SUGRA modes, this diagram
gives a 1-loop matching correction to the 4-dimensional
effective theory, which gives an operator of the form
\beq\eql{loopop}
\De\scr{L}^{(4)} \sim  \frac{1}{16\pi^2} \frac{R^{-2}}{m_*^4}
\myint d^4\th\, X^\dagger X Q^\dagger Q.
\eeq
(The appearance of the 4-dimensional Planck scale can be understood
directly by evaluating this graph in the 4-dimensional effective theory,
or by keeping track of the volume factors in the $(4 + n)$-dimensional
graph.)
% The factor of $1/(16\pi^2)$ ensures that these contributions are
% negligible even for modest hierarchy between $M_*$ and $R^{-1}$.
These operators can compete with the
anomaly-mediated contributions if the hierarchy between $M_*$ and
$R$ is modest.
For example, for $n$ extra dimensions compactified on a symmetric torus
of radius $R$, the contributions above will dominate the anomaly-mediated
contributions (discussed below) provided
\beq
M_* R \lsim \left( \frac{16\pi^2}{(2\pi)^n} \right)^{1/(2 + n)}.
\eeq
This goes to zero for large $n$, but \eg
for $n = 1$, this is $M_* R \lsim 3$, while the exponential suppression
of states with mass $M_*$ propagating between the hidden and observable
sector is $e^{\pi R M_*} \sim 10^{-4}$ (assuming the separation between
the sectors is maximal).
Although there are large uncertainties in these estimates, they do show that
models of this type are plausible.
This was invoked in \Ref{RS} as a possible mechanism to cure the problem
with negative slepton masses.%
\footnote{Note that the loop graph is finite and calculable
because of the `point splitting' due
to the spatial separation of the hidden and observable sectors,
and therefore it is an issue whether it has the right sign.}
In this paper we will investigate the more robust possibility that $M_* R$
is large enough that these effects can be neglected.

From the above discussion, it is clear that one can consider more
general scenarios with additional bulk fields, as long as these fields
do not have flavor-dependent couplings.
It may be difficult to eliminate dangerous flavor-changing
bulk fields in realistic string models.
Our focus is on the phenomenology of anomaly-mediation, and
we will not attempt to address this question here.

With these results we can write the 4-dimensional effective theory below the
scale $1/R$.
Neglecting exponentially small effects and operators of the form
\Eq{loopop}, the effective lagrangian can be written
\beq\eql{truly}
\scr{L}^{(4)}_{\rm eff} = \scr{L}_{\rm obs} + \scr{L}_{\rm hid}
+ \scr{L}_{\rm SUGRA} + O(\ep / m_*),
\eeq
where $\scr{L}_{\rm obs}$ ($\scr{L}_{\rm hid}$)
contains the observable (hidden) fields and covariant couplings
to SUGRA, and $\scr{L}_{\rm SUGRA}$ contains the SUGRA kinetic terms
and self-couplings.
Higher-dimension couplings connecting the hidden and observable
sectors are suppressed by a small parameter $\ep$ compared to the
na\"\i vely-expected suppression $1/m_*$.
The form of this lagrangian is guaranteed by the higher-dimensional
scenario discussed above, but may occur more generally.
\Eq{truly} captures what \Ref{RS} refer to as a `sequestered' sector.

% ----------------------------------------------------------------------------
\subsection{Supergravity and the Conformal Compensator\label{SUGRAsec}}
From now on, our discussion will be in the context of the 4-dimensional
effective theory given by \Eq{truly}.
We now review some aspects of 4-dimensional
$\scr{N} = 1$ SUGRA that are necessary to
understand anomaly-mediation.

In the minimal set of auxiliary fields for $\scr{N} = 1$ SUGRA,
the SUGRA multiplet contains the fields \cite{auxSUGRA}
\beq\eql{SUGRA}
e_m{}^\mu,\ 
\psi_{\mu \al},\ 
R_\mu,\ 
H,
\eeq
where $e_m{}^\mu$ is the tetrad, $\psi_{\mu\al}$ is the gravitino,
$R_\mu$ is a real auxiliary field and $H$ is a complex auxiliary field.
We will be interested in SUGRA backgrounds corresponding to flat space
and broken SUSY.
In this case, the only nonzero VEV's of the SUGRA multiplet are
\beq
\avg{e_m{}^\mu} = \de_m{}^\mu,
\qquad
\avg{H} = m_{3/2}.
\eeq
The fact that $\avg{H}$ is equal to the gravitino mass comes from
the fermionic terms, which we do not discuss here.
Note also that the usual Weyl
rescaling of the gravitational fields is not necessary
since we work in a fixed background.
In this background, we can keep track of the auxiliary field $H$
by introducing a chiral field
\beq\eql{compdef}
\Si = 1 + \th^2 H
\eeq
with conformal weight $+1$ and demanding conformal invariance of the
action \cite{GS,tensor}.%
\footnote{For another approach to SUGRA, see \Ref{WZSUGRA}.}
It is a non-trivial result that \Eq{compdef} preserves a local
super-Poincar\'e symmetry.
The field $\Si$ is called the conformal compensator, and
acts as a spurion field for the breaking of conformal invariance.
For example, if we assign all matter and gauge fields to have vanishing
conformal weight, the action for matter and gauge fields in a SUGRA
background can be written
\beq\eql{theL}
\bal
\scr{L}_{\rm matt} &= \myint d^2\th d^2\bar\th\, \Si^\dagger \Si\,
K(\Phi^\dagger e^V, \Phi)
+ \left[ \myint d^2\th\, \frac{1}{2 g^2} \tr W^\al W_\al + \hc \right]
\\
&\qquad
+\, \left[
\myint d^2\th\, \Si^3\, W(\Phi) + \hc \right]
+ \hbox{derivative\ terms}.
\eal\eeq
Note that $\Si$ does not appear in the gauge kinetic term because
the conformal weight of $W^\al W_\al$ is 3.
We can define fields with arbitrary conformal weight by rescaling by
powers of $\Si$;
for example, the chiral fields $\Phi' = \Si \Phi$ have conformal
weight $+1$.
To obtain the SUGRA scalar potential from this expression, one
must integrate out the auxiliary field $H$, with a constant term in the
superpotential adjusted to cancel the vacuum energy.

% ----------------------------------------------------------------------------
\subsection{Anomaly Mediation}
We now consider an observable sector coupled to a `sequestered' sector
by SUGRA, as discussed above.
The observable sector may have higher-dimension $M_*$-suppressed
operators coupling observable fields, but these do not mediate
SUSY breaking and can therefore be ignored.
The observable sector is therefore well-approximated by a
renormalizable theory, and we can write
\beq\bal
\scr{L}_{\rm obs} &= \myint d^4\th\,
\Si^\dagger \Si\,
Z_a Q_a^\dagger e^V Q^a
\\
&\qquad
+ \myint d^2\th\, \frac{1}{2 g^2} \tr W^\al W_\al + \hc
\\
&\qquad
+ \myint d^2\th\, \Si^3 \left(
\ka_a Q^a + \frac{1}{2} m_{ab} Q^a Q^a
+ \frac{1}{3!} \la_{abc} Q^a Q^b Q^c \right)
+ \hc
\eal\eeq
A constant term in the superpotential is required to cancel the
cosmological constant, but this can be regarded as part of
$\scr{L}_{\rm hid}$.

Consider the case where the observable sector contains no dimensionful
interactions, so $\ka_a = 0$, $m_{ab} = 0$.
In terms of the rescaled fields
\beq
Q'^a = \Si Q^a
\eeq
the lagrangian can be written
\beq\bal
\scr{L}_{\rm obs} &= \myint d^4\th\,
Z_a Q'^\dagger_a e^V Q'^a
\\
&\qquad
+ \myint d^2\th\, \frac{1}{2 g^2} \tr W^\al W_\al + \hc
\\
&\qquad
+ \myint d^2\th \left(
\frac{1}{3!} \la_{abc} Q'^a Q'^b Q'^c \right)
+ \hc
\eal\eeq
All dependence on the conformal compensator $\Si$
has completely disappeared, so there
is no SUSY breaking in the observable sector at the classical level.
It is clear that the absence of SUSY breaking is closely connected
with the conformal symmetry of the classical action that allows us to
scale away the $\Si$ dependence.
It should therefore not be surprising that the quantum conformal anomaly
gives rise to $\Si$ dependence.

This can be made precise in a number of different ways \cite{RS,GLMR}.
We will give a non-perturbative argument based on the 1PI definition
of soft mass terms \cite{ALR,GLMR}.
The origin of the $\Si$ dependence in the quantum theory is the fact
that the regulator necessarily introduces a mass scale that breaks
conformal symmetry explicitly, and therefore introduces $\Si$ dependence.
(Note that we are considering quantum effects in a fixed SUGRA background,
so we need not regulate SUGRA itself.)
$\Si$ is a spurion for conformal symmetry (with conformal weight $+1$),
and so the dependence on $\Si$ is determined by dimensional analysis.
This allows us to directly read off the $\Si$ dependence in the 1PI
effective action.
For example, the 1PI 2-point function for a chiral field $Q$ can
be written
\beq
\Ga_{\rm 1PI} = \myint d^4 x\!\! \myint d^4\th\,
Q^\dagger_a \zeta_a(\Yfund (\Si^\dagger \Si)^{-1}) Q^a
+ \cdots,
\eeq
where the dependence of $\zeta$ on $\Yfund = \partial^\mu \partial_\mu$
is a manifestation of the conformal anomaly, and the $\Si$ dependence
is determined by the fact that $\Si$ is a spurion with conformal weight $+1$.
The $F$ terms of $\Si$ give rise to a soft mass term for the scalar
component of $Q^a$, and we obtain
\beq\eql{msoftdefn}
m^2_{a}(\mu) &= \left[ \ln \zeta_a(-\mu^2 (\Si^\dagger \Si)^{-1})
\right]_{\th^2 \bar{\th}^2}
\\
\eql{msoftresult}
&= -\frac{1}{4} \frac{d\ga_a}{d\ln\mu} m_{3/2}^2,
\eeq
where $\ga_a = d\ln Z_a / d\ln\mu$ and
\beq\eql{Zdefn}
Z_a(\mu) = \zeta_a(\mu^2).
\eeq
\Eqs{msoftdefn} and \eq{Zdefn} are the definitions of the 1PI running soft
mass parameter and wavefunction factor in a superfield coupling scheme
\cite{ALR}, and \Eq{msoftresult} follows simply by differentiation.
This argument is non-perturbative, and shows that
\Eq{msoftresult} is an RG-invariant
relation that holds at all scales.
This means that the anomalous dimensions that determine the soft scalar
masses are themselves completely determined by the effective theory at the weak
scale, with no dependence on the underlying fundamental theory!

Similar arguments can be given for gaugino masses \cite{RS,GLMR}
and $A$ terms \cite{GLMR} and we obtain
\beq
m_\la = \frac{\be(g^2)}{2 g^2} m_{3/2},
\qquad
B_{a} = \frac{1}{2} \ga_a m_{3/2},
\eeq
where $\be(g^2) = d g^2 / d\ln\mu$ and $B_{a}$ determines the $A$
terms via
\beq
\De\scr{L} = \frac{1}{3!} \la_{abc} (B_{a} + B_{b} + B_{c}) 
\tilde{Q}^a \tilde{Q}^b \tilde{Q}^c + \hc,
\eeq
where $\tilde{Q}$ are the scalar components of $Q$.

% ----------------------------------------------------------------------------
\section{Slepton Masses}
% ----------------------------------------------------------------------------
We now consider the problem of negative slepton masses.
The signs of the soft masses are determined by the signs of anomalous
dimensions.
Schematically,
the anomalous dimension for a chiral field has 1-loop contributions
\beq
\ga \sim \frac{1}{16\pi^2} \left( -\la^2 + g^2 \right).
\eeq
The scalar masses therefore have signs
\beq\eql{msigns}
m^2 &\sim -m_{3/2}^2 \left( \frac{\partial\ga}{\partial\la} \be_\la
+ \frac{\partial\ga}{\partial g} \be_g \right) 
\nonumber\\
&\sim +\left( \frac{m_{3/2}}{16\pi^2} \right)^2 \left[
+\la(\la^3 - \la g^2) - g (\pm g^3) \right],
\eeq
where $\be_g \sim \pm g^3 / (16\pi^2)$.
We see that if we neglect the effects of the Yukawa couplings, an
asymptotically free gauge group gives a positive scalar mass-squared,
while gauge groups that are not asymptotically free
give a negative mass-squared.
Since the lepton Yukawa couplings are small in the MSSM, and the leptons are
charged only under the non-asymptotically free groups
$SU(2)_W \times U(1)_Y$, the slepton masses are negative.
The squark masses obtain large positive contributions from $SU(3)_C$,
and are not problematic.

In the following Subsections,
we will explore extensions of the MSSM that can give
positive slepton masses.
In the next Section, we will address electroweak symmetry breaking.

% ----------------------------------------------------------------------------
\subsection{New Lepton Yukawa Couplings\label{newYuksec}}
Perhaps the simplest way to obtain positive slepton masses
is to extend the MSSM to include
new Yukawa couplings involving leptons.
This requires new fields beyond those present in the MSSM.
(The new Yukawa couplings must be order 1 for \emph{all} lepton fields
in order to overcome the negative gauge contribution to the slepton
masses.
We therefore cannot make use of the $R$-parity violating
terms allowed in the MSSM.)
A simple possibility is to introduce an additional pair
of Higgs doublets $H'_u$ and $H'_d$ with superpotential couplings
\beq\eql{newlYuk}
\De W = (y'_e)_{jk} L_j H'_d \bar{E}_k.
\eeq
% (We will discuss other possibilities below.)
Leptoquarks are another simple possibility, but we will focus
on extra Higgs doublets.
The fields $H'_{u,d}$ must have masses at or below the electroweak
scale in order to contribute to the anomalous dimensions of the
lepton fields at the weak scale and change the sign of the slepton
masses.

There are a number of issues that arise immediately when we consider
Yukawa couplings of this form:
lepton flavor violation,
mixing between $H'_d$ and $H_d$,
and
gauge coupling unification.
We now address each of these in turn.

We first consider FCNC's.
Without special assumptions about the flavor structure, the Yukawa
couplings will not be diagonal in the basis that diagonalizes the lepton
mass matrix.
This will give a tree-level contribution to $\mu^\pm \to e^\pm e^+ e^-$
from exchange of the scalar components of $H'_d$.
At scales below the mass of the new field, this can be
parameterized by the effective interaction
\beq
\scr{L}_{\rm int}
\sim \frac{y'^2_e}{m_{H'd}^2} (\bar{\mu} e) (\bar{e} e) + \hc
\eeq
The experimental limit 
$\Ga(\mu^\pm \to e^\pm e^+ e^-) / \Ga_{\rm tot} < 1.0 \times 10^{-12}$
\cite{PDG} gives a bound
\beq[lfbound]
m_{H'd} \gsim y'_e \cdot (100\TeV).
\eeq
(The process $\mu^\pm \to e^\pm \ga$ gives somewhat weaker bounds.)
%
% The new Yukawa interaction will necessarily contribute to
% $\mu\to e\ga$ through the 1-loop graph shown in Fig.~?.
% This graph will be flavor-diagonal in a lepton basis that diagonalizes
% the new Yukawa couplings (we assume there is such a basis), and therefore
% gives rise to an effective interaction of the form
% \beq
% \scr{L}_{\rm int} \sim \frac{e y^2}{16\pi^2 M^2}\, F_{\al\be}
% \left[ m_\mu \bar{e} \si^{\al\be} \mu
% + \bar{e} \ga^\al \partial^\be \mu \right]
% + \hc,
% \eeq
% where $y$ is the Yukawa coupling and $M$ is the mass of the heaviest
% particle in the loop.
% The factor of $m_\mu$ in the first term arises from the necessity of a
% helicity flip in the amplitude.
% The derivative in the second term contributes of order $m_\mu$ to
% the amplitude, so these terms are of the same order.
% From the bound $\Ga(\mu\to e\ga) / \Ga_{\rm tot} < 4.9 \times 10^{-11}$
% \cite{PDG}
% we obtain $M \gsim y \cdot (18\TeV)$.
%
% The bound \Eq{lfbound} is very stringent because $m_{H'd}$ must be
% below the slepton mass $m_{\tilde{\ell}}$
% in order for the Yukawa term to control the
% anomalous dimensions of the sleptons.
% Since the slepton mass is $m_{\tilde\ell}^2 \sim y'^4_e \ms^2$, we obtain
% \beq
% \ms \gsim \frac{100 \TeV}{y'_e}.
% \eeq
% For $y'_e \sim 1$, this necessitates fine-tuning of order
% $M_W^2 / \ms^2 \sim 10^{-6}$ to obtain electroweak symmetry
% breaking of the correct magnitude.
Since $y'_e$ must be of order 1, this forces the soft mass for $H'_d$
to unnaturally large values that can destabilize the potential
for the other Higgs fields.

We are therefore forced to assume that there is additional
nontrivial flavor structure in the lepton sector.
Perhaps the simplest possibility is to assume that there is a
% $(Z_2)_e \times (Z_2)_\mu \times (Z_2)_\tau$ lepton
$(Z_2)^3$
flavor symmetry that forces the lepton Yukawa couplings
to be diagonal.%
\footnote{The symmetry may also be approximate, but we take it
to be exact for simplicity.}
It may appear that such symmetry forbids neutrino mixing,
but this need not be the case.
Currently, the most convincing evidence for neutrino mixing comes
from the solar and atmospheric neutrino anomalies, which are purely
disappearance effects.
These can be explained by mixing with sterile neutrinos that
carry lepton family numbers.

Even mixing among `active' neutrino flavors does not preclude the
existence of approximate lepton family number conservation for
charged leptons, which can emerge as accidental symmetries.
For example, assume that the $(Z_2)^3$
% $(Z_2)_e \times (Z_2)_\mu \times (Z_2)_\tau$ lepton
flavor symmetry is broken
by a spurion $S$ with total lepton number $+2$ in a model with
right-handed neutrinos.
(For example, $S$ may be proportional to VEV's of fields with
lepton number $+2$.)
This gives a Majorana mass for the right-handed neutrinos:
\beq
W = y_{e j} L_j H_d \bar{E}_j + y_{\nu j} L_j H_u \bar{N}_j
+ \sfrac{1}{2} M S_{jk} \bar{N}_j \bar{N}_k.
\eeq
We assume that $M \gg M_W$.
Below the scale $M$, this gives an effective interaction
\beq
W_{\rm eff} = y_{e j} L_j H_d \bar{E}_j
-\frac{y_{\nu j} y_{\nu k}}{2 M} (S^{-1})_{jk} 
(L_j H_u) (L_k H_u)
+ \cdots
\eeq
The $1/M$ term gives small Majorana neutrino masses of order
$y_\nu^2 v^2 / M$; this is the standard see-saw mechanism.
The low-energy theory contains no modification of the diagonal form
of the charged lepton Yukawa matrices because these conserve
lepton number.

% We now turn to unification.
% Adding an extra pair of doublets spoils the one-step
% gauge coupling unification observed in the MSSM.
% The only natural way to restore unification is to add
% color triplets $T, \bar{T}$
% so that the additional matter content is ${\bf 5} \oplus \bar{\bf 5}$
% of $SU(5)$.
% The fermionic components of the new doublets and triplets must have
% weak-scale masses to avoid gross conflict with experiment.
% We cannot add `$\mu$ terms' of the form
% $\mu' H'_u H'_d + \mu_T T \bar{T}$
% to the superpotential, since this would generate large `$B \mu$ terms'
% and give rise to large VEV's for some components of $H'_{u,d}$, $T$, and
% $\bar{T}$.
% This issue is closely related to the $\mu$ problem for $H_{u,d}$,
% and will be addressed below.
% Unfortunately, we will find that 

We now turn to the issue of $H'_d$--$H_d$ mixing.
This turns out to be the most severe constraint on this scenario.
Because $H'_d$ has identical quantum numbers as $H_d$ and both fields
couple to leptons, these fields will mix at one loop.
Because of this, the anomaly-mediated soft masses contain off-diagonal
terms
\beq
\De V_{\rm soft} = \De m^2 (H^\dagger_{d} H'_d + \hc),
\eeq
with $\De m^2 \sim \ms^2 y_\tau y'^3_e$.
These terms are dangerous because they give a
tadpole for $H'_d$, resulting in
$\avg{H'_d} / \avg{H_d} \sim y_\tau / y'_e$.
This in turn gives a
contribution to the electron mass
\beq
\De m_e \sim y'_e \avg{H'_d} \sim m_\tau.
\eeq
The observed value of the electron mass can be recovered only at the
expense of an unnatural fine-tuning of order
$m_e / m_\tau \sim 3 \times 10^{-4}$.
This is clearly unacceptable.

We can solve this problem in a manner consistent with 't Hooft
naturalness by adding 3 new Higgs fields
$H'_{d j}$, $H'_{u j}$ ($j = 1,2,3$).
The $(Z_2)^3$
% $(Z_2)_e \times (Z_2)_\mu \times (Z_2)_\tau$
symmetry allows us to write the superpotential couplings
\beq
W = y_{ej} L_j H_d \bar{E}_j + y'_{e jk} L_j H'_{d k} \bar{E}_j.
\eeq
In order to suppress Higgs mixing we assume the existence of an
\emph{approximate} $(Z_2)'^3$ 
% $(Z_2)'_{e} \times (Z_2)'_{\mu} \times (Z_2)'_{\tau}$
symmetry, where under $(Z_2)'_1$ (for example)
$L_1$, $H'_{d 1}$, and $H'_{u 1}$ are odd and all other fields are even.%
\footnote{Note that this allows a `$\mu$ term' $H'_{u1} H'_{d1}$.}
If the symmetry were exact, it would force the couplings $y'_{e jk}$
to be exactly diagonal.
This symmetry is violated by the electron Yukawa coupling
(since $H_d$ and $\bar{E}_1$ are even under $(Z_2)'_1$).
There are 2-loop effects proportional to $y_e^2 y'^2_e$ that contribute
to mixing between \eg\ $H'_{d1}$ and $H'_{d2}$, and so we naturally
have
\beq
y'_{e jk} = y'_{e j} \de^{\vphantom\prime}_{jk}
+ O(y_{e j} y_{e k}),
\qquad
y'_{e j} \sim 1.
\eeq
With these assumptions, the mixing between \eg\ $H'_{d1}$ and $H_d$ is
controlled by $y_{e1}$ at one loop, and so does not give an unnaturally
large contribution to the electron mass.

It may appear that the assumption of approximate lepton flavor
symmetries is somewhat artificial.
However, the small observed values of the lepton Yukawa couplings
are only natural if approximate flavor symmetries are present.
They may arise as accidental symmetries, or due to a hierarchical
breaking of flavor symmetries in a more fundamental theory.

A significant drawback of this scenario is that the only natural way to
preserve one-step gauge coupling unification is to add 3 color
triplets, so that we are adding 3 ${\bf 5} \oplus \bar{\bf 5}$'s of
$SU(5)$.
But then the 1-loop $SU(3)_C$ beta function vanishes, and
squark masses are too small.
(In fact, the only complete $SU(5)$ multiplets we can add compatible
with $SU(3)_C$ asymptotic freedom are a complete generation or
1 or 2 ${\bf 5} \oplus \bar{\bf 5}$.)
%
% \footnote{%
% In fact, when we consider electroweak symmetry breaking,
% we find that even if we add two ${\bf 5} \oplus \bar{\bf 5}$'s,
% obtaining sufficiently large squark masses requires additional
% fine-tuning.}
%
% \footnote{For example, one can consider coupling $H_{d1}$ dominantly
% to the first lepton generation, and $H_{d2}$ to the second and third
% generations.
% This gives a correction to the muon mass of order $m_\tau$, which is
% about 10 times too large.
% We therefore require only a $10\%$ fine-tuning to obtain the
% observed value of $m_\mu$.})
%
Of course, this does not mean that the model is incompatible with the
general idea of unification, since one can have intermediate
thresholds and/or non-minimal GUT groups.
However, we do give up a simple explanation of the
striking fact that the simplest possible one-step
unification appears to work in the MSSM.

% ----------------------------------------------------------------------------
\subsection{Compositeness at the Weak Scale}
\newcommand{\lc}{\La_{\rm comp}}
\newcommand{\gc}{g_{\rm comp}}
We have seen that adding new lepton Yukawa couplings requires special
flavor structure in the lepton sector.
Since gauge interactions naturally conserve flavor, it is natural
to consider the possibility that gauge interactions give rise to
positive contributions to slepton masses.
The difficulty is that this requires the 
gauge group to be asymptotically free, and hence non-Abelian.

One possibility is that the leptons are composite with a compositeness
scale $\lc$ near $\ms$.
% In this context, `compositeness' means
That is, we assume
that there is a new
asymptotically-free gauge group that gets strong at the scale $\lc$,
whose non-perturbative dynamics produces light fermions with the quantum
numbers of the observed leptons.
% (Of course, other light fermions and scalars might also be composite.)
The scale $\lc$ must be low enough so that the non-Abelian gauge group
dominates the anomalous dimensions of the lepton degrees of freedom
at the scale where their masses are generated.
This scenario therefore requires lepton compositeness near the weak
scale.

This scenario is severely constrained by searches for deviations in
cross-sections at colliders (`compositeness searches' \cite{PDG}).
In the effective theory below $\lc$, we expect 4-fermion operators
of the form
\beq
\scr{L}_{\rm int} \sim \frac{(4\pi)^2}{\lc^2} (\bar{e} e)^2.
\eeq
The factor of $(4\pi)^2$ is inserted based on `\naive dimensional
analysis' \cite{NDA}:
$\lc$ is defined as the scale at which the underlying theory becomes
strongly interacting in the sense that loop corrections are unsuppressed,
and this should coincide with the scale at which the effective theory
becomes strong.
With 4-fermion couplings of this strength, current experimental limits
give \cite{PDG}
\beq\eql{lcbound}
\lc \gsim 18\TeV.
\eeq
This bound does not necessarily translate into a large value
for $\ms$, because the gauge coupling $\gc$ can be large at the
scale where the slepton masses are generated.
To estimate this, we use the 1-loop RG equations for
$\gc$,
\beq
\mu \frac{d}{d\mu} \left( \frac{1}{\gc^2} \right)
= \frac{b}{8\pi^2},
\eeq
and define $\lc$ to be the scale where $\gc^2 \sim 16\pi^2 / b$.
Demanding that the slepton mass $m_{\tilde\ell} \sim \gc^2 \ms$ is
larger than $\lc$ then gives
\beq
\ms \gsim \frac{\lc}{g^2(m_{\tilde\ell})}
\sim \frac{b}{8\pi^2} m_{\tilde\ell} \left(
\frac{1}{2} + \ln\frac{m_{\tilde\ell}}{\lc} \right).
\eeq
For example, for $\lc \sim m_{\tilde\ell} \sim 20\TeV$, $b \sim 5$
we obtain $\ms \sim 600\GeV$.
There are clearly large theoretical uncertainties in these estimates,
but they show that this scenario is not impossible.
(Note that such heavy sleptons do not give large naturalness-spoiling
contributions to the Higgs mass because the lepton Yukawa couplings are
small.)

One theoretical puzzle that arises in this scenario is the question of
why the compositeness scale should be so close to $\ms$, since these
scales have a different origin.
(In the scenario of \Ref{RS}, the SUSY breaking sector
that sets the scale $M_{\rm SUSY}$ is literally in another world!)
However the near coincidence of these scales may be natural because
the scalars and gauginos charged under the
compositeness gauge group get masses of order
$m_{\tilde\ell} \sim g_{\rm comp}^2 \ms$.
The beta function of the compositeness gauge group is more negative
below this scale.
Then the running coupling $g_{\rm comp}$ may therefore become strong
rapidly, naturally explaining why $\lc$ is close to $\ms$.
This mechanism will be especially efficient if the compositeness
gauge coupling is `walking' above the compositeness scale.

% The effective theory below this scale is a non-supersymmetric theory
% that contains only fermions charged under the compositeness gauge
% group.
% The beta function of the compositeness gauge group is made more
% negative by the fact that there are fewer particles charged under it,
% and the running coupling $g_{\rm comp}$ may rapidly become strong
% below the scale $g_{\rm comp}^2 \ms$.

% The viability of this scenario is enhanced if the compositeness gauge
% coupling is `walking' \cite{walk}.
% This means that above the scale $m_{\tilde\ell}$, the running
% coupling $g_{\rm comp}$ is close to its critical value, but is running
% slowly.
% In that case, the scale $g_{\rm comp}^2 \ms$ may be significantly
% above $\ms$, and the theory will become strong rapidly
% below $g_{\rm comp}^2 \ms$.

Having composite leptons and elementary quarks may be difficult
to reconcile with one-step gauge coupling unification.
The simplest way to ensure gauge coupling unification is to
construct a model in which quarks as well as leptons are
composite, and that the `preons' as well as the composites
occur in complete $SU(5)$ multiplets.

Finding a model that embodies this mechanism is clearly
non-trivial.
In particular, the compositeness dynamics is inherently
non-supersymmetric below the scale $m_{\tilde\ell}$,
and therefore not under theoretical control.
However, we regard this scenario as an interesting (albeit
speculative) possibility.

% This class of models will likely be further constrained in the near future by
% an improved bound on $\mu\to e\ga$ \cite{MEGA}.

% ----------------------------------------------------------------------------
\subsection{Horizontal Gauge Symmetries}
Another possible way to couple leptons to a non-Abelian gauge symmetry
is to assume that lepton flavor symmetry is gauged.
% However, this scenario suffers from a host of problems, and we do not
% consider it very attractive.
Since the gauge group must be unbroken near the weak scale (in order
for its anomalous dimensions to affect the slepton masses), this
forces us to build a model of flavor at the weak scale.
% Since the flavor gauge group contains generators that change
% the flavor, it is very hard to see how one can avoid flavor-changing
% neutral currents.
This theory contains gauge bosons that change flavor quantum numbers,
and therefore does not naturally suppress FCNC's.
For example, the process $\mu^\pm \to e^\pm e^+ e^-$
gives a bound on the mass of the horizontal gauge boson
$M_{\rm hor} \gsim 85\TeV$
(assuming a horizontal gauge coupling of order $g_2$ and no alignment).
This requires $\ms$ to be of comparable size, and necessitates
fine-tuning in order to obtain electroweak symmetry breaking.%
\footnote{This would be somewhat ameliorated if the gauge coupling
of the horizontal gauge group were larger.
However, in the limit where the gauge coupling approaches the
perturbative limit, the model becomes a composite model that
violates flavor.}
We will not consider this possibility further.

% ----------------------------------------------------------------------------
\section{Electroweak Symmetry Breaking}
% ----------------------------------------------------------------------------
We now turn to electroweak symmetry breaking.
One crucial issue is the $\mu$ term.
A tree-level $\mu$ term in the observable sector can be written
\beq
\De\scr{L} = \myint d^2\th\, \Si^3 \mu H_u H_d + \hc
\eeq
where $\Si$ is the conformal compensator discussed in Subsection
\ref{SUGRAsec}.
After scaling out the compensator, we see that
$B\mu = \mu m_{3/2} \sim 16\pi^2 \ms \mu$, which is too large
since other soft masses are of order $\ms$.
This is a direct result of having a dimensionful parameter in the
observable sector, which breaks conformal invariance at tree level
and ruins the anomaly-mediation result that all soft masses are
loop suppressed.
An attempt to use the Giudice-Masiero mechanism \cite{GM} requires
extra interactions between the hidden and observable sector, and does
not give $\mu^2 \sim B\mu \sim \ms^2$ unless parameters are
adjusted to special values.
Since the original `$\mu$ problem' is precisely to explain why an
arbitrary parameter (a tree-level $\mu$ term in the MSSM) has a
special value ($\sim M_W$) we do not regard this as an attractive
possibility.

Clearly the most natural possibility is that an effective $\mu$ term
is generated by SUSY breaking at the weak scale.
We are led to the idea that the $\mu$ term arises as the VEV
of a singlet field $S$ with a superpotential coupling
$S H_u H_d$.
If the scalar mass-squared term for $S$ is negative, this will naturally
give $\avg{S} \sim \ms$.
The idea that the $\mu$ term arises from the VEV of a singlet at the
weak scale has been considered by many authors \cite{NMSSM}.
We will also assume that the slepton mass problem is solved by the
addition of 3 extra pairs of Higgs doublets $H'_{u j}, H'_{d j}$
($j = 1,2,3$) and triplets $T, \bar{T}$, as discussed in
Subsection~\ref{newYuksec}.
We also add 3 additional singlets $R_j$ ($j = 1,2,3$),
which we take to be odd under the corresponding factor of the
approximate $(Z_2)'^3$
% $(Z_2)'_e \times (Z_2)'_\mu \times (Z_2)'_\tau$ symmetry.
For simplicity, we also assume that the $(Z_2)^3$
% $(Z_2)_e \times (Z_2)_\mu \times (Z_2)_\tau$
symmetry discussed in Subsection~\ref{newYuksec} above is exact.
The most general renormalizable superpotential invariant under the
(exact and approximate) flavor symmetries is then
\beq\eql{thespot}
\bal
W &= \la_H S H_d H_u + \frac{\ka}{3} S^3
+ \la_T S T \bar{T} + \frac{b_j}{2} S R_j^2
\\
&\qquad
+ %\sum_{j = 1}^3 \left[
\la'_{H j} S H'_{d j} H'_{u j} 
+ a_{u j} R_j H'_{d j} H_u + a_{d j} R_j H_d H'_{u j}
+ (y'_e)_j H'_{d j} L_j \bar{E}_j % \right]
\\
&\qquad
+ (y_u)_{jk} Q_j H_u \bar{U}_k
+ (y_d)_{jk} H_d Q_j \bar{D}_k.
\eal\eeq
The conventional lepton Yukawa couplings
\beq
\De W = y_{e j} H_d L_j \bar{E}_j
\eeq
are invariant under % $(Z_2)_e \times (Z_2)_\mu \times (Z_2)_\tau$,
$(Z_2)^3$, but break the approximate $(Z_2)'^3$
% $(Z_2)'_e \times (Z_2)'_\mu \times (Z_2)'_\tau$
symmetry, so we expect additional terms in the superpotential
suppressed by factors proportional to lepton Yukawa couplings.
In particular, there are 1- and 2-loop diagrams that give rise to
mixings: $H'_{d j}$--$H_{d}$, $H'_{u j}$--$H_{u}$, and $R_j$--$S$.
% (see Fig.~2).
These will give rise to mixing masses, \eg
\beq
\De V_{\rm soft} = \De m^2_{S R j} (S^\dagger R_j + \hc)
\eeq
with
\beq
\De m^2_{S R j} \sim \ms^2 a_{u j} y'_{e j} y_{e j} \la_H.
\eeq
This will give rise to a tadpole for $R_j$;
if $m_R^2 \sim \ms^2$ is positive, we have
$\avg{R_j} \sim a_{u j} y'_{e j} y_{e j} \la_H \avg{S}$.
These effects are all suppressed by small lepton Yukawa couplings,
and can therefore be neglected for purposes of discussing electroweak
symmetry breaking.

% \begin{figure}[t]
% \centerline{\epsfxsize=5.0in\epsfbox{mix.eps}}
% \smallskip
% \caption{Fig.~2.\ \ Loop graphs that mix Higgs fields.  Note that
% all graphs are proportional to small lepton Yukawa couplings $y_e$.}
% \end{figure}
% 

The soft SUSY breaking terms are completely determined by the
dimensionless couplings in the superpotential, together with the
gauge couplings.
(Explicit formulas are given in the Appendix.)
We look for a solution with
\beq
\avg{S} \sim \avg{H_u} \sim \avg{H_d} \sim M_W,
\eeq
with all other VEV's small.
(As discussed above, if the soft masses of $H'_{u,d}$ and $R$ are
positive and of order $\ms$, they will get small VEV's suppressed
by lepton Yukawa couplings.)
% We need to avoid $\avg{H'_d} \ne 0$ because otherwise the leptons will
% get large masses, since $y'_e \sim 1$ is required to get positive
% slepton masses.
% We also need to avoid $\avg{H'_u} \ne 0$, because otherwise there will
% be a tadpole $\avg{S} \avg{H'_u} H'_d$ that will give $\avg{H'_d} \ne 0$.
The role of the various terms in the superpotential \Eq{thespot}
are as follows:

$\bullet$
The triplets $T, \bar{T}$ are required because the term
$S T \bar{T}$ gives an important negative contribution to the
mass-squared of $S$ from the term (see \Eq{msigns})
\beq
m_S^2 = \ms^2 \la_T^2(15 \la_T^2 - 16 g_3^2) + \cdots.
\eeq
This is required in order for $\avg{S}$ to be sufficiently large.
(The term $S H_u H_d$ gives a tadpole for $S$, but results in
small values for $\avg{S}$, and therefore a small $\mu$ term.)

% Eduardo's version
$\bullet$
The singlets $R_j$ are required to obtain electroweak symmetry
breaking together with the observed value of the top quark mass,
as we now explain.
The $SU(2)$ and $U(1)$ gauge couplings give negative contributions to
both $m^2_{Hu}$ and $m^2_{Hd}$ that can induce electroweak
symmetry breaking.
In addition, $m^2_{Hu}$ receives a
contribution from the top yukawa coupling
\beq\eql{mHucontrol}
m_{Hu}^2 = 18 y_t^2 (y_t^2 - \sfrac{8}{9} g_3^2) \ms^2 + \cdots.
\eeq
In the small $\tan\be$ regime where $y_b, y_\tau \ll 1$, $y_t$ is the
only parameter that differentiates between $m^2_{Hu}$ and $m^2_{Hd}$.
Therefore, $\tan\be$ is completely determined by $y_t$ and so
the top mass is determined by $y_t$ alone.
We find that it is not possible to get a top mass above $\sim 145\GeV$,
as illustrated in Fig.~2.
For $y_t \lsim g_3$, the contribution in \Eq{mHucontrol} is negative
and therefore $\tan\be > 1$.
For $y_t \gsim g_3$ we obtain $\tan\be < 1$, and as $y_t$ increases
both $\tan\be$ and $m_t$ decrease until the electroweak
symmetry breaking is lost.
The introduction of the singlets $R_j$ with couplings
$R H_u H'_d$ and $R H_d H'_u$ give additional contributions to
$m^2_{Hu}$ and $m^2_{Hd}$,
thus eliminating the correlation between $y_t$ and $\tan\be$ just
described. 

\begin{figure}[t]
\centerline{\epsfxsize=5.0in\epsfbox{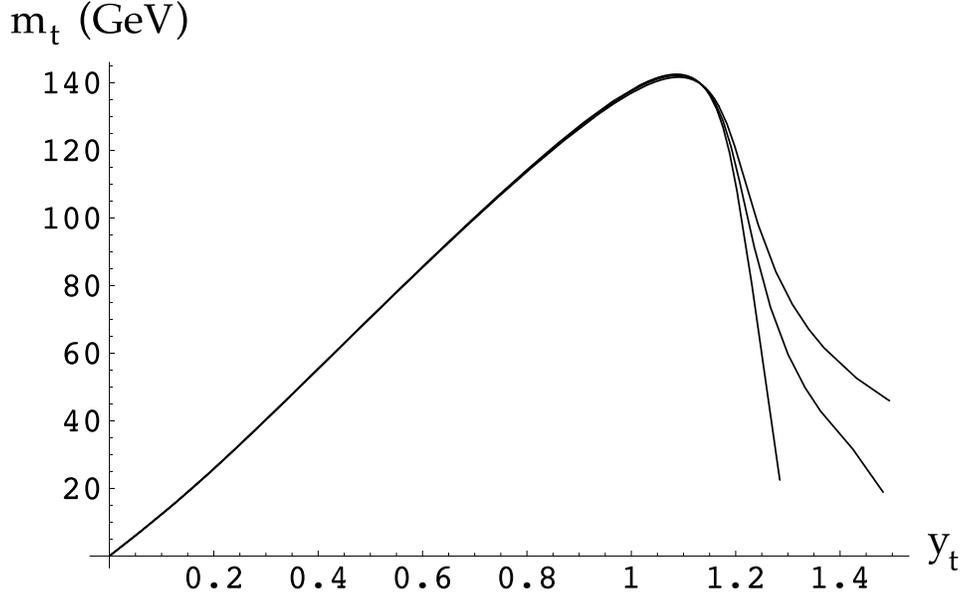}}
\smallskip
\caption{Fig.~2.\ \ Top quark mass as a function of $y_t$
for different values of $\la_T$.
From the upper curve to the lower, the values of $\la_T$
are = $0.35$, $0.25$, and $0.15$.
The curves end at a finite value of $y_t$
because the model no longer breaks electroweak symmetry.
The other parameters are $\la_H = 0.15$,
$\la_T = 0.15$,
$\la_{H}' = 0.15$,
$\ka = 0.3$,
$a_d = 0$,
$a_u = 0$,
$b = 0$,
$y'_e = 0.95$,
$y_t = 1$.
$y_b = 0.1$
$y_\tau = 0.05$
}
\end{figure}

The above discussion assumes that $\tan\be \sim 1$.
We also explored the possibility of large $\tan\be$ in the
model without the singlets $R_j$, but find solutions
only with large fine-tuning in several parameters.
The reason for this is that $m_{Hu}^2 /\ms^2$ is bounded from below
by the anomalous dimension formulas.
This means that the only way to obtain large $\tan\be$ is to have
small $B\mu/\ms^2$, and hence small $\mu/\ms$.
But since there is a chargino mass of order $\mu$, this means that
$\ms \gg v$, which can only be obtained by extreme fine-tuning.
We find solutions only
with fine tuning $\sim 10^{-3}$ in both $y_t$ and $y_b$.

The couplings $R H_u H'_d$ and $R H_d H'_u$ have another purpose,
besides allowing us to reproduce the observed top mass.
Even if we could somehow live with $m_t \simeq 145\GeV$, we would not
have a viable solution without the singlet $R$.
The problem is that for `generic' choices of the parameters, the
Higgs VEV is $v \gsim 5\ms$, where $v = 174\GeV$ is required to
reproduce the correct values of $M_W$ and $M_Z$.
This results in a small value of $\ms$, and hence many superpartner
masses are too small.
This problem can be cured only by adjusting parameters so that the
potential is near the critical point for electroweak symmetry
breaking (so that $v \sim \ms$).
However, this is not possible without the couplings $a_u$ and $a_d$.

This explains why we need the couplings
$R H_u H'_d$ and $R H_d H'_u$: they allow us to control
$m_{Hu}^2$ and $m_{Hd}^2$ independently, eliminating the correlation
between $y_t$ and $\tan\be$ discussed above.
We must introduce new singlets $R_j$ for this purpose rather than using
a term $S H_u H'_d$ because we must avoid a tadpole
$\avg{S} \avg{H_u} H'_d$ that would
otherwise give a large VEV for $H'_d$.

% The reason for this is that natural solutions are only possible if all
% mass parameters are of order 1 or larger in units of $\ms$.
% However, $\mu \gsim 1$ (in units of $\ms$) requires $B\mu \gsim 1$;
% in order to obtain electroweak symmetry breaking with large $\tan\be$
% we then require $m_{Hd}^2$ large and positive and $m_{Hu}^2$
% large and negative.
% But this not possible, since $m_{Hu}^2$ is bounded from below
% by the anomalous dimension formulas.
% The only solutions we find have $\mu \ll 1$, so that $\ms \gg v$,
% and have large fine-tuning in both $y_t$ and $y_b$.
% (For example, we find solutions with sufficiently large $\mu$, but
% with fine tuning $\sim 10^{-3}$ in $y_t$ and $2 \times 10^{-3}$ in $y_b$.)

We now discuss some important features of the superpotential \Eq{thespot}
that allow this model to be realistic.

$\bullet$
The terms proportional to $\la_H$, $\la_{H}'$, $\la_T$, and $b$
give rise to effective `$\mu$ terms' that give masses proportional
to $\avg{S}$ for the fermion fields of
$H_{u,d}$, $H'_{u,d}$, $T$, $\bar{T}$, and $R$.
% All of these fields carry standard-model gauge quantum numbers,
% and so these masses must be at the weak scale.
We therefore need all of these couplings to be nonzero.

% $\bullet$
% The coupling $S R^2$ is required to give masses to
% the fermion components of $R$.
% Because $R$ is a singlet that does not couple directly to observable
% fields, this mass may be significantly below the weak scale.

$\bullet$
As already discussed above, $y_t$ and $\la_T$ give
important negative contributions to $m^2_{Hu}$ and $m^2_S$,
respectively (see \Eq{msigns}),
which allow $\avg{H_{u,d}}, \avg{S}$ to have realistic values.

% The top Yukawa coupling gives an essential negative contribution to
% $m_{Hu}^2$ proportional to $-y_t^2 g_3^2$ (see \Eq{msigns} or
% the detailed formulas in the Appendix).
% This can be viewed as a manifestation of radiative electroweak symmetry
% breaking in this class of models.
% 
% $\bullet$
% The coupling $\la_T S T \bar{T}$ gives an important negative contribution
% to $m_S^2$ proportional to $-\la_T^2 g_3^2$, similar to the negative
% contribution to $m_{Hu}^2$ from $y_t$.
% This is essential to obtain a sufficiently large value for $\avg{S}$.

$\bullet$
The couplings $S H'_u H'_d$, $R H_u H'_d$, and $R H_d H'_u$
give a positive contribution
to $m_{H'u}^2$ and $m_{H'd}^2$, necessary to obtain
$\avg{H'_u} \simeq \avg{H'_d} \simeq 0$.

$\bullet$
The coupling $\ka$ is required to break a $U(1)$ Peccei--Quinn symmetry
that would otherwise give rise to a weak-scale axion.

$\bullet$
R-parity can be extended to the new fields above by taking the scalar
components of $S$, $R$, $H'_{u,d}$, $T$, and $\bar{T}$ to have
R-parity $+1$, while their fermion components have R-parity $-1$.
This means that this model is of the conventional R-parity conserving
type, with a stable LSP.

$\bullet$
As discussed so far, the fields $T, \bar{T}$ carry an exactly
conserved quantum number, so the lightest particle carrying this
quantum number is stable.
The constraints on such particles from direct experiment
\cite{CDFcolorstable} and cosmology \cite{colorstablecosmo}
are quite stringent.
However, these are easily avoided by small Yukawa couplings
that allow the $T$ to decay:
\beq
\De W = \ep_j R_j T \bar{D}_j.
\eeq
These violate flavor symmetries, but \eg $\ep_1 \sim 10^{-5}$
is suffucient for a $T$ particle with mass $\sim 150\GeV$
to have decay with $c\tau \sim 1\ {\rm mm}$.
This is safe from the constraints, while still being consistent
with the existence of the approximate flavor symmetries discussed
above.

One unattractive feature of this model is that some degree of
fine-tuning is required to avoid light superpartner masses.
% We believe this problem is generic to anomaly-mediated models.
As already discussed above, this arises because for
`generic' values of the parameters, we have $v \gsim 5 \ms$,
resulting in a small value for $\ms$ and hence light superpartner
masses.
We require
$\ms \gsim 100\GeV$, which means that we need $v/\ms \lsim 1$.
%
% The most stringent bound on $\ms$ comes from the $SU(2)_W$
% gaugino mass:
% \beq\eql{mwtilde}
% m_{\tilde{W}} = (1 + n_2) g_2^2 \ms,
% \eeq
% where $n_2$ is the number of extra doublets
% compared to the MSSM ($n_2 = 3$ in the present model).
% \Eq{mwtilde} is generically significantly
% smaller than other soft masses because the
% $SU(2)_W$ beta function is accidentally small.%
%
% \footnote{%
% One can try to make other anomalous dimensions smaller by choosing
% small values of the superpotential couplings,
% but this leads to other difficulties, such as unwanted light states.
% One can contemplate increasing the $SU(2)_W$ beta function by increasing
% $n_5$, but $SU(3)_C$ loses asymptotic freedom for $n_5 = 3$, resulting in
% small (or negative) squark masses.
% Of course we could simply give up gauge coupling unification and add many
% additional doublets, but we do not pursue this here.}
%
% (The 1-loop $SU(2)_W$ beta function would be zero in the MSSM without
% the Higgs doublets).
%

The reason for the `generically large' Higgs VEV is that
the quartic terms in the Higgs potential are small.
The coefficient of the quartic term arising from the $SU(2)_W \times U(1)_Y$
$D$ terms is $\frac{1}{8}(g_1^2 + g_2^2) \simeq 0.066$;
there is also a quartic term proportional to $\la_H^2$ from the $F$ terms,
but $\la_H$ cannot be $\sim 1$ % too much larger than $\simeq 0.3$
because this would give a too-large positive contribution to $m_S^2$,
resulting in $\avg{S} \simeq 0$.
We must therefore tune the parameters so that the Higgs potential is near the
critical point where $\avg{H_u} = \avg{H_d} = 0$, so that $v$
is small in units of $\ms$.
Near the critical point, the Higgs VEV is determined by a formula
of the form $v^2 \sim m^2 / \la$ where $m^2$ is the coefficient of a
quadratic term and $\la$ is the coefficient of a quartic term.%
\footnote{The potential also has cubic terms, but they are less
important near the critical point.}
Since $m^2$ is analytic in the couplings, near the critical point
\beq
v \sim (c - c_{\rm crit})^{1/2},
\eeq
where $c$ is a coupling that acts as the control parameter.
% This explains the fact that $v$ varies rapidly near the
% critical point. % , as seen in Fig.~4.
This is illustrated in Fig.~3.

\begin{figure}[t]
\centerline{\epsfxsize=5.0in\epsfbox{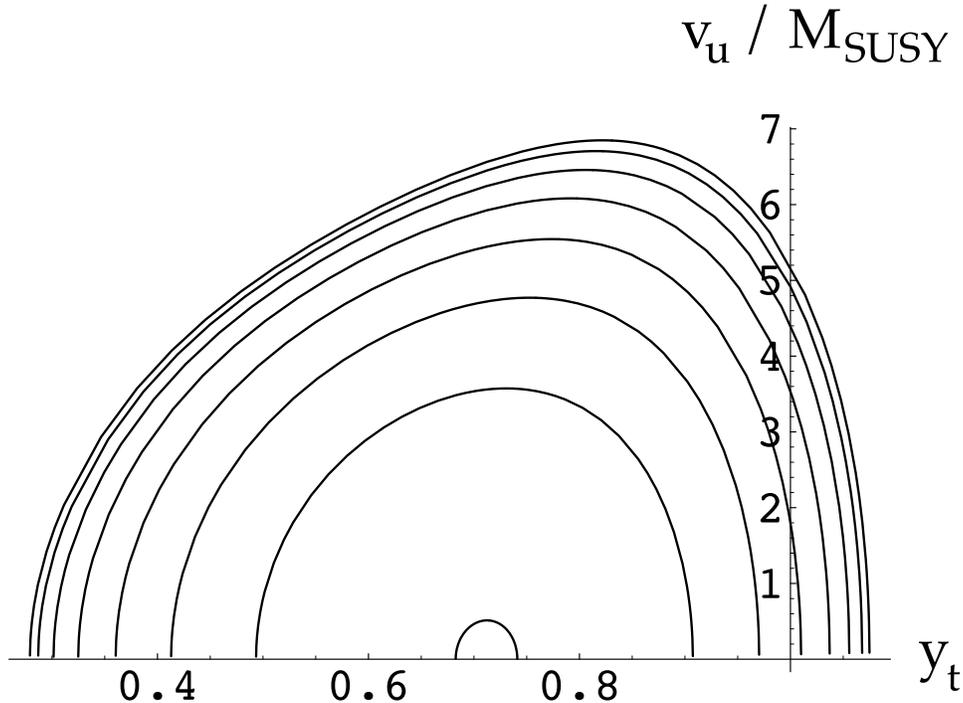}}
\smallskip
\caption{Fig.~3.\ \ Fine-tuning of electroweak symmetry breaking.
The plot shows $v_u$ as a function of $y_t$ for values of
$a_{u 3}$ ranging from $0.1$ (upper curve) to $0.8$
(lower curve).
The values of the other coupling constants are as in
the example in Section 4.}
\end{figure}

Because the tree-level quartic terms in the Higgs potential are small,
loop corrections to these terms can be important.
In addition to the usual $y_t^4$ corrections of the MSSM,
there are large $a_{u,d}^4$ corrections arising from the Yukawa terms
$a_u R H_u H'_d$ and $a_d R H_d H'_u$.
These generate non-supersymmetric corrections to the quartic terms in
the Higgs potential
\beq
\De V = \De\la_u (H_u^\dagger H_u)^2
+ \De \la_d (H_d^\dagger H_d)^2,
\eeq
with
\beq\bal
\De \la_u &\simeq \frac{1}{8\pi^2} \left[
3 y_t^4 \ln\frac{m_{\tilde{t}}}{m_t}
+ 3 a_u^4 \ln \frac{m_{R}}{m_{\tilde{R}}} \right],
\\
\De \la_d &\simeq \frac{1}{8\pi^2} 
3 a_d^4 \ln \frac{m_{R}}{m_{\tilde{R}}},
\eal\eeq
where $R$ ($\tilde{R}$) is the scalar (fermion) component of the
superfield $R$.
(We have taken universal values for $a_{u j}, a_{d j}$ for simplicity.)
These corrections increase the mass of the neutral Higgs boson, and
also reduce the amount of fine-tuning required for a realistic
solution.

We have not attempted to explore the full parameter space of
this model, but we have found regions of the parameter space with
realistic solutions.
% Small mixing effects suppressed by loop factors will not greatly
% affect our results.
The solutions we find have $\avg{S} \sim 7\ms$ and small $\la_H$;
this gives a sufficiently large $\mu$ term
while avoiding large positive
contributions to $m_S^2$, $m_{Hu}^2$, and $m_{Hd}^2$ from
$\la_H$.
This requires small values of $\ka$, and so we are driven to a regime
where the scalar component of $S$ is light
(it is a pseudo-Nambu--Goldstone boson of the Peccei--Quinn symmetry).
This means that the Higgs sector differs significantly from the MSSM
Higgs sector, and the lightest \CP-even Higgs boson
% with significant couplings to electroweak gauge bosons
can be heavier than $M_Z$
even without the radiative corrections discussed above.
Also, because of the presence of extra neutral fermions and
moderate $\mu$ terms, the neutralino-chargino
degeneracy discussed in \Refs{RS,GLMR} is not realized in our
model.
We have checked that all scalar mass-squared terms are positive
at these points.

As an example, we take
$\la_H = 0.15$,
$\la_T = 0.17$,
$\la_{H}' = 0.15$,
$\ka = 0.075$,
$a_d = 1$,
$a_u = 0.5$,
$b = 0.05$,
$y'_e = 0.95$,
$y_t = 1$.
(We take universal values for $\la'_{H j}, a_{u j}, a_{d j},
b_j$, and set $y_b = y_\tau = 0$.)
In this case, we find
\beq
\frac{v_u}{\ms} = 1.0,
\quad
\frac{v_d}{\ms} = 0.34,
\quad
\frac{\avg{S}}{\ms} = 6.9.
\eeq
This gives $\ms = 160\GeV$, and reproduces the observed top quark mass
with $\tan\be = 3.1$.
For this solution, the lightest neutralino is mainly $R$
fermion, and
chargino and neutralino are not degenerate:
$m_{\tilde{\chi}_1^0} = 87\GeV$,
$m_{\tilde{\chi}_1^\pm} = 129\GeV$.
Near degeneracy of the lightest neutralino and chargino is a
generic feature of an anomaly-mediated spectrum in the MSSM
provided there are no additional neutral fermions and the
$\mu$ term is sufficiently large \cite{RS,GLMR}.
In the present case, both of these conditions are violated
($\mu = 160\GeV$).

The lightest neutral \CP-even Higgs has mass $91\GeV$,
and contains a significant admixture of $S$ scalar.
The lightest neutral \CP-odd Higgs has mass $180\GeV$
and is mainly $S$ scalar.
The sleptons have masses ranging from $130\GeV$ (for sneutrinos)
to $310\GeV$ (for right-handed selectrons),
while squarks of the first two generations have masses near
$430\GeV$.
The gluino mass is $380\GeV$.
There are also colored bosons from the fields
$T, \bar{T}$ with mass $170\GeV$.
Of course, this point is only meant as an example,
but it shows that the sparticle spectrum can be
quite conventional apart from the large number of additional
states compared to the MSSM.

We now discuss the fine tuning.
The most sensitive dependence of $M_Z$ on fundamental
parameters is%
\footnote{
This is the measure of fine-tuning proposed in \Ref{BG}.
It has been emphasized that this measures sensitivity, which
does not imply fine-tuning if the sensitivity is high for all
{\it a priori} allowed parameters \cite{AC}.
In the present case, for generic values of the input parameters
the sensitivity is much less (see Fig.~3),
and so the large sensitivity is a sign of fine-tuning.}
\beq
\frac{\partial \ln M_Z}{\partial\ln y_t} = 77,
\qquad
\frac{\partial \ln M_Z}{\partial\ln a_u} = 13,
\eeq
Here, $y_t$ is the running parameters at the weak scale.
If we express the fine-tuning in terms of fundamental parameters at
a higher scale, the fine-tuning is significantly less due to the
infrared quasi-fixed point for $y_t$ \cite{iryt}.
For example, the sensitivity is reduced by a factor of 10 in terms
of $y_t(\mu \sim 10^8\GeV)$.

Since the parameters we have chosen are somewhat 
fine-tuned in terms of parameters at the weak scale,
the VEV's may be sensitive to loop corrections.
However, the existence of a critical point where 
$\avg{H_{u,d}} \to 0$ should survive the inclusion
of loop effects, and there will be realistic solutions when the
parameters are adjusted to appropriate values.
We believe that our analysis is therefore sufficient to conclude that
the model can be realistic.

% ----------------------------------------------------------------------------
\section{Conclusions}
% ----------------------------------------------------------------------------
We have investigated the possibility that supersymmetry breaking
is communicated to the observable sector by the
recently-discovered mechanism of anomaly mediation \cite{RS,GLMR}.
This is automatically the leading mechanism if the lagrangian of the
world has the `sequestered' form
\beq
\scr{L} = \scr{L}_{\rm obs} + \scr{L}_{\rm hid} + \scr{L}_{\rm SUGRA}
+ O(\ep / m_*),
\eeq
where $\scr{L}_{\rm obs}$ ($\scr{L}_{\rm hid}$) contains only the
observable (hidden) sector fields and their couplings to supergravity,
and $\scr{L}_{\rm SUGRA}$ contains the supergravity kinetic terms and
self-interactions.
Higher-dimension operators that directly connect the hidden and
observable sectors are suppressed by an additional small parameter
$\ep$ compared to the na\"\i vely-expected suppression $1/m_*$,
where $m_*$ is the reduced Planck scale.
Such a mechanism is naturally realized if the hidden sector is
a parallel universe in higher dimensions \cite{RS}.
In that case $\ep \sim e^{-M_* R}$, where $M_*$ is the higher-dimensional
Planck scale and $R$ is the size of the `large' extra dimensions.

This mechanism automatically solves the SUSY flavor problem, and is
also theoretically very appealing.
However, it suffers from several glaring phenomenological problems:
slepton masses are negative, and the $\mu$ problem is more difficult
to solve.
Previous discussions have assumed additional direct couplings between
the hidden and observable sector, requiring special choices of
parameters to ensure that all SUSY breaking is at the weak scale.
In this paper, we have constructed a realistic model with \emph{no}
direct couplings between the hidden and observable sector.
This means that \emph{all}
dimensionful parameters in the observable sector arise from anomaly
mediation, and are therefore determined by
anomalous dimensions at the weak scale.
In this type of model, the superpartner masses are directly
determined by dimensionless couplings of fields at the weak scale;
both the masses and couplings are experimentally accessible,
so these relations are in principle testable.
These models also generally contain a large number of additional
charged particles near the weak scale that are subject to
experimental study.
Our model extends the minimal supersymmetric standard model
by 3 pairs of Higgs doublets, 1 vector-like pair of color triplets,
and 4 new singlets.
These fields are constrained by approximate discrete flavor symmetries
in order to avoid lepton flavor-changing neutral currents and
unwanted mixing.
The $\mu$ term is generated by the vacuum expectation value of a singlet.
The model is tightly constrained, and it is non-trivial that the
tree-level and loop effects can combine to give a realistic
spectrum.
This class of models provides a theoretically well-motivated and
phenomenologically interesting framework for physics beyond the
standard model.

{\bf Note Added:}
While completing this paper, we received an interesting paper by
A. Pomerol and R. Rattazzi \cite{PR}
that considers a different mechanism that may make
anomaly-mediation realistic.
We also received \Refs{FMRS,GGW}, which consider experimental
signatures of the nearly degenerate neutralino/chargino.

% ----------------------------------------------------------------------------
\section*{Acknowledgments}
% ----------------------------------------------------------------------------
We thank R. Sundrum for discussions.
This work was supported by the National
Science Foundation under grant PHY-98-02551,
and by the Alfred P. Sloan Foundation.

\newpage
% ----------------------------------------------------------------------------
\appendix{Appendix}
% ----------------------------------------------------------------------------
As discussed in the main text, we ignore mixing between
$H_{u,d}$ and $H'_{u,d}$ for simplicity, so we
look for solutions with $\avg{H'_{u,d}} = 0$.
The relevant terms in the superpotential for electroweak symmetry
breaking are then
\beq
W = \la_H S H_d H_u + \frac{\ka}{3} S^3.
\eeq
We can choose $\la_H$ and $\ka$ to be real without loss of generality.
The most general solution up to $SU(2)_W \times U(1)_Y$ rotations
can be parameterized as
\beq
\avg{H_u} = \pmatrix{0 \cr v_u \cr},
\quad
\avg{H_d} = \pmatrix{v_d \cos\th \cr v_d \sin\th \cr} e^{i\al},
\quad
\avg{S} = x e^{i\ga},
\eeq
where $v_u$, $v_d$, and $x$ are all real.
The $F$ terms in the potential are
\beq
V_F = |\la_H H_d H_u + \ka S^2|^2
+ \la_H^2 |S|^2 \left( |H_u|^2 + |H_d|^2 \right),
\eeq
which gives
\beq\eql{avf}
\bal
\avg{V_F} &= \la_H^2 v_u^2 v_d^2 \cos^2\th
+ 2 \la_H \ka x^2 v_u v_d \cos\th \cos(2\ga - \al)
+ \ka^2 x^4
\\
&\qquad
+ \la_H^2 x^2 (v_u^2 + v_d^2).
\eal
% \avg{V_F} = (\la_H v_u v_d + \ka x^2)^2 +
% \la_H^2 x^2 (v_u^2 + v_d^2).
\eeq
The relevant soft SUSY breaking terms are
\beq
\!\!\!\!\!\!\!
V_{\rm soft} = m_{Hu}^2 |H_u|^2 + m_{Hd}^2 |H_d|^2 + m_S^2 |S|^2
+ \left( \la_H A_{\la H} S H_d H_u
+ \sfrac{1}{3} \ka A_\ka S^3 + \hc \right).
\eeq
The soft terms are all real because all the couplings are real, so
this gives
\beq\eql{avsoft}
\bal
\avg{V_{\rm soft}} &=
m_{Hu}^2 v_u^2 + m_{Hd}^2 v_d^2 + m_S^2 x^2
\\
&\qquad
+ 2 \la_H A_{\la H} x v_u v_d \cos\th \cos(\ga + \al)
+ \sfrac{2}{3} \ka A_\ka x^3 \cos(3\ga).
\eal
% + 2 A_{\la H} x v_u v_d \cos(\al) + \sfrac{2}{3} A_\ka x^3 \cos(3\al).
\eeq
Finally, the $SU(2)_W \times U(1)_Y$ $D$ terms are
\beq
V_D = \sfrac{1}{8} (g_1^2 + g_2^2) (|H_u|^2 - |H_d|^2)^2
+ \sfrac{1}{2} g_2^2 |H_u^\dagger H_d|^2,
\eeq
which gives
\beq\eql{avd}
\avg{V_D} = \sfrac{1}{8} (g_1^2 + g_2^2) (v_u^2 - v_d^2)^2
+ \sfrac{1}{2} g_2^2 v_u^2 v_d^2 \sin^2\th.
\eeq
The sum of \Eqs{avf}, \eq{avsoft}, and \eq{avd} is to be minimized
with respect to $v_u$, $v_d$, and $x$.
We have checked explicitly that all scalar mass-squared terms are
positive at the solution, so we have at least a local minimum.
% All the realistic solutions we find have $\al = 0$.

We are only interested in minima that preserve $U(1)_{\rm EM}$ and
do not spontaneously violate \CP, so we will assume that
$\th = \al = \ga = 0$ from now on.

We now turn to the fermion mass matrices.
Anomaly mediation predicts the phase of the gaugino masses, and so
we must be careful about relative signs.
We define the anomaly-mediated
gaugino masses to be given by \Eq{exactsoft}.
The neutralinos
$\{ \tilde{B}, \tilde{W}_3, \tilde{H}_u^0, \tilde{H}_d^0, \tilde S \}$
have (Majorana) mass matrix
\beq
% \!\!\!\!\!\!\!\!\!
% \scr{M}_{\rm neutralino} = \pmatrix{
\scr{M} = \pmatrix{
-m_1 & 0 & g_1 v_u / \sqrt{2} & -g_1 v_d / \sqrt{2} & 0 \cr
0 & -m_2 & -g_2 v_u / \sqrt{2} & g_2 v_d / \sqrt{2} & 0 \cr
g_1 v_u / \sqrt{2} & -g_2 v_u / \sqrt{2} & 0 & \la_H x & \la_H v_d \cr
-g_1 v_d / \sqrt{2} & g_2 v_d / \sqrt{2} & \la_H x & 0 & \la_H v_u \cr
0 & 0 & \la_H v_d & \la_H v_u & 2\ka x \cr}.
\eeq
In addition, there are neutral fermions
$\{ \tilde{R}_j, \tilde{H}'_{u j}, \tilde{H}'_{d j} \}$
with mass matrix
\beq
\scr{M}_{{\rm neutralino}, j} = \pmatrix{
b_j x & -a_{d j} v_d & a_{u j} v_u \cr
-a_{d j} v_d & 0 & \la'_{H j} x \cr
a_{u j} v_u & \la'_{H j} x & 0 \cr}.
\eeq
In the basis $\{ \tilde{W}^+, \tilde{H}_u^+ \}$,
$\{ \tilde{W}^-, \tilde{H}_d^- \}$
the chargino (Dirac) mass matrix is
\beq
% \scr{M}_{\rm chargino} = 
\scr{M} = 
\pmatrix{-m_2 & g_2 v_d \cr
g_2 v_u & -\la_H x \cr}.
\eeq

We now consider the scalar mass matrices.
Write
\beq\bal
H_u &= \pmatrix{ H^+_u \cr v_u + (h^0_u + i A^0_u) / \sqrt{2} \cr},
\quad
H_d = \pmatrix{ v_d + (h^0_d + i A^0_d) / \sqrt{2} \cr H^-_d \cr},
\\
S &= (x + s + i A_S) / \sqrt{2},
\qquad\qquad\qquad
R = (r + i A_R) / \sqrt{2},
\eal\eeq
{\it etc\/}.
The mass matrix for the \CP-even neutral bosons
$\{ h_u^0, h_d^0, s \}$ is
\beq\bal
(\scr{M}^2)_{11}
&= m^2_{Hu} + \la_H^2 (v_d^2 + x^2)
+ \sfrac{1}{4} (g_1^2 + g_2^2) (3 v_u^2 - v_d)^2,
\\
(\scr{M}^2)_{12}
&= 2 \la_H^2 v_u v_d + \la_H \ka x^2 + \la_H A_{\la H} x
- \sfrac{1}{2} (g_1^2 + g_2^2) v_u v_d,
\\
(\scr{M}^2)_{13}
&= 2 \la_H^2 x v_u + 2 \la_H \ka x v_d + \la_H A_{\la H} v_d,
\\
(\scr{M}^2)_{22}
&= m_{Hd}^2 + \la_H^2 (v_u^2 + x^2)
+ \sfrac{1}{4} (g_1^2 + g_2^2) (3 v_d^2 - v_u^2),
\\
(\scr{M}^2)_{23}
&= 2 \la_H^2 x v_d + 2 \la_H \ka x v_u + \la_H A_{\la H} v_u,
\\
(\scr{M}^2)_{33}
&= m_S^2  + \la_H^2 (v_u^2 + v_d^2) + 2 \la_H \ka v_u v_d + 
6 \ka^2 x^2 + 6 \ka A_\ka x,
\eal\eeq
The \CP-even bosons $\{h'^0_{u j}, h'^0_{d j}, r_j \}$ have mass matrix
\beq\bal
(\scr{M}_j^2)_{11}
&= m^2_{H'u j} + a_{d j}^2 v_d^2 + \la'^2_{H j} x^2
+ \sfrac{1}{4} (g_1^2 + g_2^2) (v_u^2 - v_d^2),
\\
(\scr{M}_j^2)_{12}
&= (\la_H \la'_{H j} + a_{u j} a_{d j}) v_u v_d + \ka \la'_{H j} x^2 
+ \la'_{H j} A'_{H j} x,
\\
(\scr{M}_j^2)_{13}
&= (\la_H a_{d j} + \la'_{H j} a_{u j}) x v_u
+ b_j a_{d j} x v_d + a_{d j} A_{ad j} v_d,
\\
(\scr{M}_j^2)_{22}
&= m_{H'd j}^2 + a_{u j}^2 v_{u j}^2 + \la'^2_{H j} x^2
- \sfrac{1}{4} (g_1^2 + g_2^2) (v_u^2 - v_d^2),
\\
(\scr{M}_j^2)_{23}
&= \la_H a_{u j} x v_d + b_j a_{u j} x v_u + \la'_{H j} a_{d j} x v_d
+ a_{u j} A_{au j} v_u,
\\
(\scr{M}_j^2)_{33}
&= m_{R j}^2
+ b_j^2 x^2 + b_j \ka x^2 + b_j \la_H v_d v_u + a_{u j}^2 v_u^2
+ a_{d j}^2 v_d^2 + 2 b_j A_{b j} x,
\\
\eal\eeq
The \CP-odd bosons $\{ A_u^0, A_d^0, A_S \}$ have mass matrix
\beq\bal
(\scr{M}^2)_{11}
&= m^2_{Hu} + \la_H^2 (v_d^2 + x^2)
+ \sfrac{1}{4} (g_1^2 + g_2^2) (v_u^2 - v_d^2),
\\
(\scr{M}^2)_{12}
&= - \la_H \ka x^2 - \la_H A_{\la H} x,
\\
(\scr{M}^2)_{13}
&= 2 \la_H \ka x v_d - \la_H A_{\la H} v_d,
\\
(\scr{M}^2)_{22}
&= m_{Hd}^2 + \la_H^2 (v_u^2 + x^2)
- \sfrac{1}{4} (g_1^2 + g_2^2) (v_u^2 - v_d^2),
\\
(\scr{M}^2)_{23}
&= 2 \la_H \ka x v_u - \la_H A_{\la H} v_u,
\\
(\scr{M}^2)_{33}
&= m_S^2  + \la_H^2 (v_u^2 + v_d^2) - 2 \la_H \ka v_u v_d
+ 2 \ka^2 x^2
- 6 \ka A_\ka x,
\eal\eeq
This matrix has a zero eigenvalue corresponding to the Nambu--Goldstone
boson absorbed in the Higgs mechanism.
The \CP-odd bosons $\{ A'_{u j}, A'_{d j}, A_{R j} \}$ have mass matrix
\beq\bal
(\scr{M}_j^2)_{11}
&= m^2_{H'u j} + a_{d j}^2 v_d^2 + \la'^2_{H j} x^2
+ \sfrac{1}{4} (g_1^2 + g_2^2) (v_u^2 - v_d^2),
\\
(\scr{M}_j^2)_{12}
&= (a_{u j} a_{d j} - \la_H \la'_{H j}) v_u v_d
- \ka \la'_{H j} x^2 - \la'_{H j} A'_{H j} x,
\\
(\scr{M}_j^2)_{13}
&= (\la'_{H j} a_{u j} -\la_H a_{d j}) x v_u
+ b_j a_{d j} x v_d - a_{d j} A_{ad j} v_d,
\\
(\scr{M}_j^2)_{22}
&= m_{H'd j}^2 + a_{u j}^2 v_u^2 + \la'^2_{H j} x^2
- \sfrac{1}{4} (g_1^2 + g_2^2) (v_u^2 - v_d^2),
\\
(\scr{M}_j^2)_{23}
&= (\la'_{H j} a_{d j} - \la_H a_{u j}) x v_d
+ b_j a_{u j} x v_u - a_{u j} A_{au j} v_u,
\\
(\scr{M}_j^2)_{33}
&= m_{R j}^2 
+ b_j^2 x^2 - b_j \ka x^2
- \la_H b_j v_u v_d + a_{u j}^2 v_u^2 + a_{d j}^2 v_d^2
- 2 b_j A_{b j} x,
\eal\eeq

The gaugino masses are determined from the beta functions by
\beq
m_\la = m_{3/2} \frac{\be_g}{g},
\eeq
where $\be_g = dg /d\ln\mu$.
The 1-loop gauge beta functions are
\beq\bal
16\pi^2 \be_3 &= -(3 - n_3) g_3^3,
\\
16\pi^2 \be_2 &= (1 + n_2) g_2^3,
\\
16\pi^2 \be_1 &= (11 + n_2 + \sfrac{2}{3} n_3) g_1^3,
\eal\eeq
where $n_3$ is the number of pairs of vector-like triplets
$({\bf 3}, {\bf 1})_{-\frac{2}{3}} \oplus
(\bar{\bf 3}, {\bf 1})_{\frac{2}{3}}$,
and $n_2$ is the number of pairs of vector-like doublets
$({\bf 1}, {\bf 2})_{1} \oplus
({\bf 1}, {\bf 2})_{-1}$,
relative to the MSSM.
(In our model, $n_2 = 3$, $n_5 = 1$.)
The numerical values of the gauge couplings in $\overline{\rm DR}$
are \cite{FJJ}
\beq
g_1(1\TeV) = 0.363,
\quad
g_2(1\TeV) = 0.638,
\quad
g_3(1\TeV) = 1.1.
\eeq

The scalar mass-squared parameters can be easily read off from the
anomalous dimensions
\beq
\ga_a = \frac{d\ln Z_a}{d\ln\mu}
\eeq
as
\beq
m^2_a = -\frac{m_{3/2}^2}{4} \frac{d\ga_a}{d\ln\mu}
= -\frac{m_{3/2}^2}{4} \sum_g
\frac{\partial \ga_a}{\partial g} \be_g,
\eeq
where the sum runs over all couplings $g$, and
$\be_g = dg / d\ln\mu$.
The $A$ terms are given by
\beq
A_{abc} = \frac{m_{3/2}}{2} (\ga_a + \ga_b + \ga_c).
\eeq
In the present model, the 1-loop anomalous dimensions are given by
\beq\bal
16\pi^2 \ga_S 
&= -4 \la_H^2 - 6 \la_T^2
- 4 \ka^2 - \sum_{j = 1}^3 \left( 4 \la'^2_{H j} + b_j^2 \right),
\\
16\pi^2 \ga_{Hu}
&= -2 \la_H^2 - 6 y_t^2 + 3 g_2^2 + g_1^2
- 2 \sum_{j = 1}^3 a_{u j}^2,
\\
16\pi^2 \ga_{Hd}
&= -2 \la_H^2 - 6 y_b^2 - 2 y_\tau^2
+ 3 g_2^2 + g_1^2 - 2 \sum_{j = 1}^3 a_{d j}^2,
\\
16\pi^2 \ga_{H'u j}
&= -2 \la_{H j}'^2  + 3 g_2^2 + g_1^2 - 2 a_{d j}^2,
\\
16\pi^2 \ga_{H'd j}
&= -2 \la_{H j}'^2 - 2 y'^2_{e j} + 3 g_2^2 + g_1^2 - 2 a_{u j}^2,
\\
16\pi^2 \ga_T
&= -2 \la_T^2
+ \sfrac{16}{3} g_3^2 + \sfrac{4}{9} g_1^2,
\\
16\pi^2 \ga_{R j}
&= -4 a_{u j}^2 - 4 a_{d j}^2 - 2 b_j^2,
\\
16\pi^2 \ga_{tL}
&= -2 y_t^2 - 2 y_b^2 + \sfrac{16}{3} g_3^2 + 3 g_2^2 + \sfrac{1}{9} g_1^2,
\\
16\pi^2 \ga_{tR}
&= -4 y_t^2 + \sfrac{16}{3} g_3^2 + \sfrac{16}{9} g_1^2,
\\
16\pi^2 \ga_{bR}
&= -4 y_b^2 + \sfrac{16}{3} g_3^2 + \sfrac{4}{9} g_1^2,
\\
16\pi^2 \ga_{\tau L}
&= -2 y_\tau^2 - 2 y'^2_{e 3} + 3 g_2^2 + g_1^2,
\\
16\pi^2 \ga_{\tau R}
&= -4 y_\tau^2 - 4 y'^2_{e 3} + 4 g_1^2.
\eal\eeq

\newpage
% ----------------------------------------------------------------------------


\begin{references}
% ----------------------------------------------------------------------------
\bibitem{Hidden}
A.H. Chamseddine, R. Arnowitt, P. Nath, \PRL{49}{970}{1982};
R. Barbieri, S. Ferrara, C.A. Savoy, \PLB{119}{343}{1982};
L.J. Hall, J. Lykken, S. Weinberg, \PRD{27}{2359}{1983}.
For a review, see 
H.P. Nilles, \PR{110}{1}{1984}.

\bibitem{SUSYFCNC}
F. Gabbiani, E. Gabrielli, A. Masiero, L. Silvestrini, hep-ph/9604387,
\NPB{477}{321}{1996}.

\bibitem{SUSYflavor}
See \eg\ M. Leurer, Y. Nir, N. Seiberg, \NPB{420}{468}{1994};
M. Dine, R. Leigh, N. Kagan, \PRD{48}{4269}{1993};
D.B. Kaplan, M. Schmaltz, \PRD{49}{3741}{1994};
L.J. Hall, M. Murayama, \PRD{48}{4269}{1993}.

\bibitem{RS}
L. Randall and R. Sundrum, hep-th/9810155.

\bibitem{HW}
P. Ho\v rava, E. Witten, hep-th/9510209, \NPB{460}{506}{1996};
E. Witten, hep-th/9602070, \NPB{471}{135}{1996};
P. Ho\v rava, E. Witten, hep-ph/9603142, \NPB{475}{94}{1996}.

\bibitem{Dvac}
Z. Kakushadze, hep-th/9806044;
J. Lykken, E. Poppitz, S.P. Trivedi, hep-th/9806080.

\bibitem{GLMR}
G.F. Giudice, M.A. Luty, H. Murayama, R. Rattazzi, hep-ph/9810442.

\bibitem{ALR}
N. Arkani-Hamed, M.A. Luty, R. Rattazzi, hep-ph/9803290,
\PRD{58}{115005}{1998}.

\bibitem{GM}
G.F. Giudice, A. Masiero, \PLB{206}{480}{1988}.

\bibitem{auxSUGRA}
K.S. Stelle, P.C. West, \PLB{74}{330}{1978};
S. Ferrara, P. van Nieuwenhuizen, \PLB{74}{333}{1978}.

% \PLB{76}{404}{1978};
% K.S. Stelle, P.C. West, \PLB{77}{376}{1978}.

\bibitem{GS}
W. Siegel, S.J. Gates Jr., \NPB{147}{77}{1979};
S.J. Gates Jr., to appear.

\bibitem{tensor}
E. Cremmer, S. Ferrara, L. Girardello, A. Van Proeyen,
\NPB{212}{413}{1983}.

\bibitem{WZSUGRA}
J. Wess, B. Zumino, \PLB{74}{51}{1978};
\PLB{79}{394}{1978};
% R. Grimm, J. Wess, and B. Zumino \NPB{152}{255}{1979};
J.A. Bagger, \NPB{211}{302}{1983}.
For a review, see J. Wess and J.A. Bagger,
{\it Supersymmetry and Supergravity},
Princeton (1992).

\bibitem{PDG}
Particle Data Group, \EPJC{3}{1}{1998}.

\bibitem{NMSSM}
P. Fayet, \NPB{90}{104}{1975};
R.K. Kaul, P Majumdar, \NPB{199}{36}{1982};
R. Barbieri, S. Ferrara, C.A. Savoy, \PLB{119}{343}{1982};
H.P. Nilles, M. Srednicki, D. Wyler, \PLB{120}{346}{1983};
J.M. Fr\`e re, D.R.T. Jones, S. Raby, \NPB{222}{11}{1983};
J.P. Derendinger, C.A. Savoy, \NPB{237}{307}{1984};
For a phenomenological discussion, see
J. Ellis, J.F. Gunion, H.E. Haber, L. Roszkowski, F. Zwirner,
\PRD{39}{844}{1989}.

\bibitem{NDA}
A.V. Manohar, H. Georgi, \NPB{234}{189}{1984};
H. Georgi, L. Randall, \NPB{276}{241}{1986}.
For applications to supersymmetric theories, see
M.A. Luty, hep-ph/9706235, \PRD{57}{1531}{98};
A.G. Cohen, D.B. Kaplan, A.E. Nelson, hep-ph/9706275,
\PLB{412}{301}{97}.

% \bibitem{delphi}
% DELPHI collaboration, hep-ex/9903071.

\bibitem{CDFcolorstable}
CDF collaboration, hep-ex/9904010.

\bibitem{colorstablecosmo}
R.N. Mohapatra, V. Tepliz, hep-ph/9804420, \PRL{81}{3079}{1998}.

\bibitem{BG}
R. Barbieri, G.F. Giudice, \NPB{306}{63}{1988}.

\bibitem{AC}
G.W. Anderson, D.J. Casta\~ no, hep-ph/9409419,
\PLB{347}{300}{1995}.

\bibitem{iryt}
B. Pendleton, G.G. Ross, \PLB{98}{291}{1981};
C.T. Hill, \PRD{24}{691}{1981}.

% \bibitem{MNY}
% S. Mizuta, D. Ng, M. Yamaguchi, \PLB{300}{96}{1993}.

% \bibitem{CDG}
% C.H. Chen, M. Drees, J.F. Gunion, hep-ph/9512230,
% \PRL{76}{2002}{1996}.

\bibitem{PR}
A. Pomarol, R. Rattazzi, hep-ph/9903448.

\bibitem{FMRS}
J.L. Feng, T. Moroi, L. Randall, M. Strassler, S. Su,
hep-ph/9904250.

\bibitem{GGW}
T. Gherghetta, G.F. Giudice, J.D. Wells, hep-ph/9904378.

\bibitem{FJJ}
P.M. Ferreira, I. Jack, D.R.T. Jones, hep-ph/9605440,
\PLB{387}{80}{1996}.

\end{references}
\end{document}